\documentclass[%
reprint,
amsmath,amssymb,
aps,
superscriptaddress,
prc
]{revtex4-1}

\usepackage{graphicx}
\usepackage{dcolumn}
\usepackage{bm}
\usepackage{hyperref}

\usepackage[capitalise]{cleveref}
\usepackage{comment}
\usepackage{color}

\begin{document}
	
	
	\title{Studies of beauty hadron and non-prompt charm hadron production in pp collisions at $\sqrt{s}$=13 TeV within a transport model approach}
	\author{Jialin He}
	\affiliation{School of Mathematics and Physics, China University of
		Geosciences (Wuhan), Wuhan 430074, China}
	\author{Xinye Peng}
	\affiliation{School of Mathematics and Physics, China University of
		Geosciences (Wuhan), Wuhan 430074, China}
	\affiliation{Key Laboratory of Quark and Lepton Physics (MOE) and Institute
		of Particle Physics, Central China Normal University, Wuhan 430079, China}
	\author{Xiaoming Zhang}
	\affiliation{Key Laboratory of Quark and Lepton Physics (MOE) and Institute
	of Particle Physics, Central China Normal University, Wuhan 430079, China}
	\author{Liang Zheng}\email{zhengliang@cug.edu.cn}
	\affiliation{School of Mathematics and Physics, China University of
	Geosciences (Wuhan), Wuhan 430074, China}
	\affiliation{Shanghai Research Center for Theoretical Nuclear Physics, NSFC and Fudan University, Shanghai 200438, China}
	\affiliation{Key Laboratory of Quark and Lepton Physics (MOE) and Institute
	of Particle Physics, Central China Normal University, Wuhan 430079, China}

	\date{\today}
	
	\begin{abstract}

In high-energy proton proton ($pp$) collisions at the LHC, non-prompt charm hadrons, originating from beauty hadron decays, provide a valuable probe for beauty quark dynamics, particularly at low transverse momentum where direct beauty measurements are challenging. We employ A Multi-Phase Transport Model (AMPT) of string melting version coupled with PYTHIA8 initial conditions to study the beauty hadron and non-prompt charm hadron productions in $pp$ collisions at $\sqrt{s} = 13$ TeV. In this work, the beauty quark mass during the generation stage has been increased to reproduce the measured $b\bar{b}$ cross section, and a beauty flavor specific coalescence parameter $r_{BM}^b$ is introduced to match LHCb measurements of beauty baryon to meson ratios. With these refinements, AMPT achieves a reasonable agreement with experimental data on beauty hadron yields and non-prompt charm hadron production from ALICE and LHCb. We present the transverse momentum and multiplicity dependence of non-prompt to prompt charm hadron ratios, providing new insights into the interplay between beauty quark production and hadronization process. We emphasize that the multiplicity dependence of the non-prompt to prompt charm hadron productions can be useful to constrain the flavor dependences of the coalescence dynamics. This work establishes a unified framework for future studies of heavy quark transport and collective flow behavior in small collision systems. 
    
	\end{abstract}
	

	\maketitle
	
	
	\section{Introduction}
	\label{sec:level1}

Measurements of hadrons containing charm or beauty quarks in proton proton ($pp$) collisions are essential for testing Quantum Chromodynamics (QCD) calculations~\cite{Andronic:2015wma}. Due to their large masses, heavy quarks are predominantly produced in the initial stages of the collision via hard partonic scatterings, making their inclusive production cross sections calculable within perturbative QCD using a factorization framework\cite{Collins:1989gx,Apolinario:2022vzg,Bedjidian:2004gd,Schweda:2014tya}. State of the art calculations, including FONLL~\cite{Cacciari:2005rk}, GM-VFNS~\cite{Kniehl:2004fy,Helenius:2018uul}, and next-to-next-to-leading order (NNLO) QCD~\cite{Mangano:1991jk}, generally provide a good description of experimental data by factorizing the process into parton distribution functions, perturbative partonic cross sections, and non-perturbative fragmentation functions.

However, recent experimental results have issued a significant challenge to the assumption of universal hadronization for heavy quarks, which implies that a heavy flavor parton fragments into hadrons independently of the collision system. A growing body of evidence, particularly the enhancement of heavy flavor baryon to meson ratios in $pp$ collisions compared to electron positron ($e^+e^-$) collisions at the LHC energy, suggests that the hadronization process of heavy flavor particles is sensitive to the surrounding environment~\cite{ALICE:2023sgl,ALICE:2025wrq,Altmann:2024kwx}.
Several complementary theoretical approaches have been proposed to explain this observed enhancement. Statistical hadronization models assume the thermal production of hadrons at chemical freeze-out and incorporate an augmented spectrum of heavy flavor baryon states, which qualitatively describe the system size dependence of heavy flavor baryon production with the transition from a canonical ensemble in small systems to a grand-canonical ensemble in heavy-ion collisions~\cite{Chen:2020drg,Dai:2024vjy,He:2019vgs,He:2019tik}. A recent comprehensive comparison of heavy-quark recombination models in heavy-ion collisions~\cite{Zhao:2023nrz} highlights the importance of hadronization dynamics in shaping final-state observables. It has been conceived that the high energy $pp$ collisions can create a sufficiently dense parton system in which heavy quarks may undergo scatterings within the short-lived evolving parton medium and hadronize by recombining with nearby light quarks via a coalescence mechanism, thereby favoring baryon formation relative to mesons~\cite{Minissale:2020bif,Minissale:2024gxx,Zhao:2024ecc}. Furthermore, modified string fragmentation models, incorporating color reconnection beyond the leading color approximation to include junction topologies, also improve descriptions of enhanced baryon production~\cite{Bierlich:2014xba,Bierlich:2015rha,Bierlich:2023okq}. Within this broader context, A Multi-Phase Transport Model (AMPT), which dynamically simulate partonic scattering and coalescence hadronization for heavy quarks, can provide a unified framework to investigate interplay between the bulk medium and heavy flavor hadronization~\cite{Zhang:2025pqu,Zheng:2024xyv,Zheng:2019alz,Lin:2021mdn}.

Direct measurements of beauty hadrons are experimentally challenging, especially at low transverse momentum ($p_T$). An effective way to overcome this limitation is through non-prompt charm hadrons, which originate from beauty hadron decays. These particles provide an indirect but powerful probe of beauty quark production and hadronization, extending experimental sensitivity to lower $p_T$~\cite{ALICE:2023brx,ALICE:2021mgk,ALICE:2024xln}. 
Recent measurements reveal that the non-prompt $\Lambda_c/D^0$ ratio is significantly larger in $pp$ collisions than in $e^+e^-$ collisions, and follows a trend similar to that of prompt charm baryons. This observation suggests the presence of a common underlying mechanism for baryon formation across charm and beauty flavors~\cite{ALICE:2023wbx}. To further disentangle contributions from initial heavy-quark production and from hadronization dynamics, it is essential to investigate these non-prompt to prompt charm hadron ratios for both meson and baryon states as a function of event multiplicity~\cite{ALICE:2023brx}. The multiplicity dependence of non-prompt charm production provides further insight into the interplay between hard and soft processes, including multi-parton interactions and possible final state effects in small systems. The AMPT model combines the dynamical parton evolution system with a flavor dependent spatial coalescence model, providing unique opportunities to link microscopic transport dynamics with the statistical descriptions of hadronization.

In this work, we employ the AMPT model with PYTHIA8 initial conditions to study beauty quark dynamics via non-prompt charm hadron production in $\sqrt{s}=13$ TeV $pp$ collisions. The string melting version of AMPT model incorporating a partonic cascade and coalescence hadronization, has successfully described baryon to meson enhancements and collective flow in small systems~\cite{Zhang:2025pqu,Zheng:2024xyv}. We provide a systematic strategy to achieve reasonable descriptions to the $b\bar{b}$ cross section at 13 TeV~\cite{Bai:2024pxk} and beauty baryon yield at forward rapidity~\cite{LHCb:2023wbo} in the AMPT framework. Using this improved model, we then analyze the $p_T$ and multiplicity dependence of beauty decayed non-prompt charm hadron to prompt charm hadron ratios. This allows us to disentangle the underlying mechanisms driving the observed multiplicity trends in experiments. This approach not only provides a quantitative baseline to exclusively interpret the heavy flavor hadron productions in small systems but also offers a foundation for future studies of heavy quark evolution, shedding a light to understand multiplicity dependence in the transition from dilute to dense QCD matter.


The rest of this paper is organized as follows. Sect.~\ref{sec:formalism} provides a detailed introduction to the AMPT model and explains the modifications made to its key parameters. Sect.~\ref{sec:results} presents the primary results, starting with beauty hadron production in Sect.~\ref{subsec:beauty} and non-prompt c-hadron production in Sect.~\ref{subsec:nonprompt}. This is followed by an investigation into the multiplicity dependence of the non-prompt to prompt hadron ratio in Sect.~\ref{subsec:nptop}. Finally, Sect.~\ref{sec:Summary} summarizes the findings of this study.
    
\begin{figure*}[htbp!]
	\begin{center}
		\includegraphics[width=1.0\textwidth]{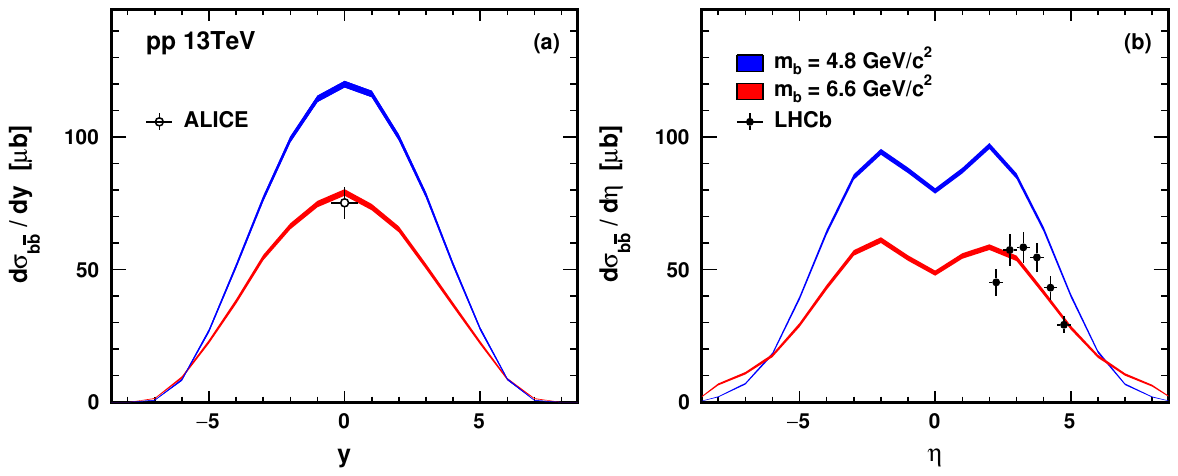}
		\caption{$y$ (left) and $\eta$ (right) distribution of the $b\bar{b}$ production cross section in $pp$ collisions at $\sqrt{s}=13$~TeV. The red and blue curves represent results generated by PYTHIA8 with the $b$ quark effective mass $m_b=4.8$ and $m_b=6.6$~GeV$/c^2$, respectively. Experimental data from ALICE~\cite{ALICE:2024xln} and LHCb~\cite{LHCb:2016qpe} are shown in black open and solid markers for $y$ and $\eta$, respectively.} 
		\label{fig:bbbar}
	\end{center}
\end{figure*}

\begin{figure}[htbp]
	\begin{center}
		\includegraphics[width=0.5\textwidth]{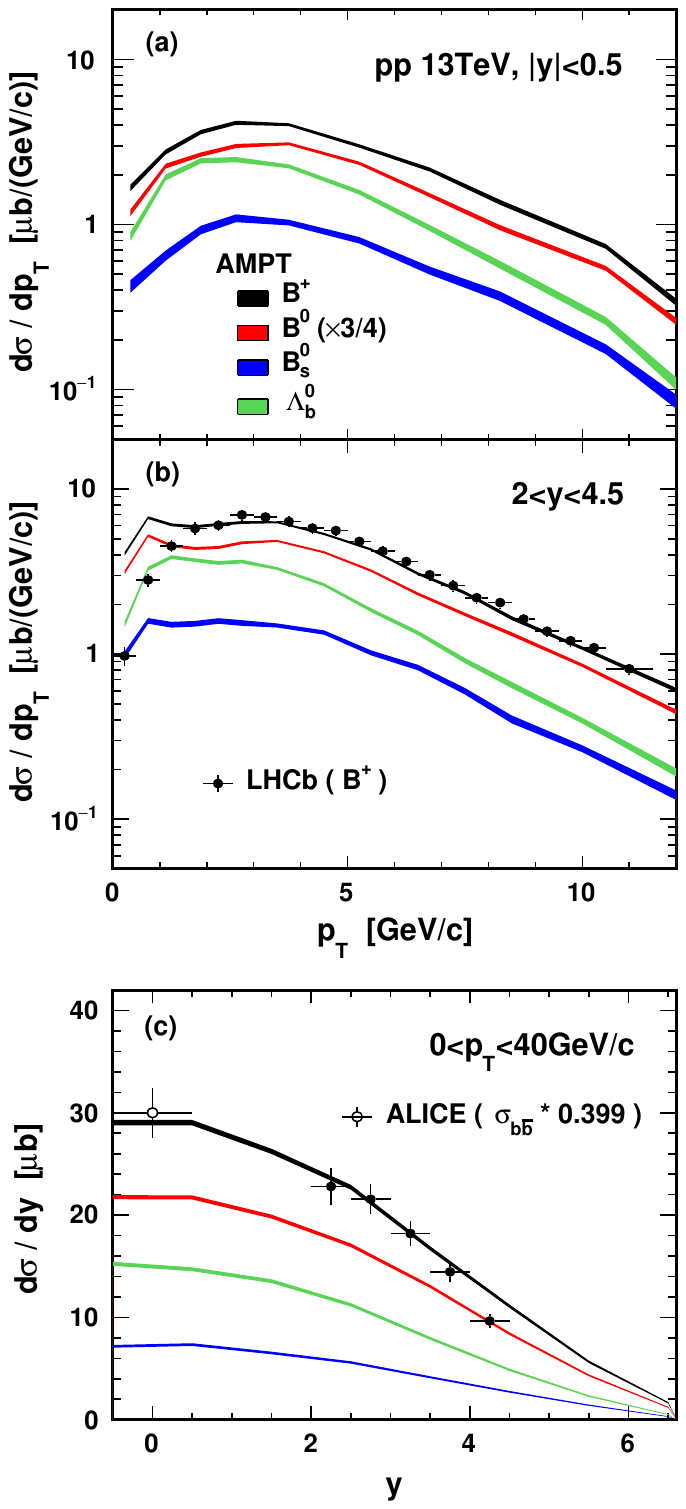}
		\caption{The $p_{T}$-differential cross section of beauty hadrons in $pp$ collisions at $\sqrt{s}=13$~TeV on mid-rapidity ($|y|<0.5$, panel (a)) and forward rapidity ($2<y<4.5$, panel (b)), and as a function of $y$ for $0<p_T<40$~GeV$/c$ (panel (c)). The black, red, blue, and green lines show AMPT results for $B^0$, $B^+$, $B_s$, and $\Lambda_b^0$, respectively. Solid and open black markers indicate $B^+$ results from LHCb~\cite{LHCb:2017vec} and ALICE~\cite{ALICE:2024xln}. The ALICE data are obtained by scaling the measured $b\bar{b}$ cross section with a fragmentation fraction of 0.399~\cite{DELPHI:2011aa}.}
		\label{fig:bhadrons}
	\end{center}
\end{figure}

\begin{figure*}[htbp]
	\begin{center}
		\includegraphics[width=0.88\textwidth]{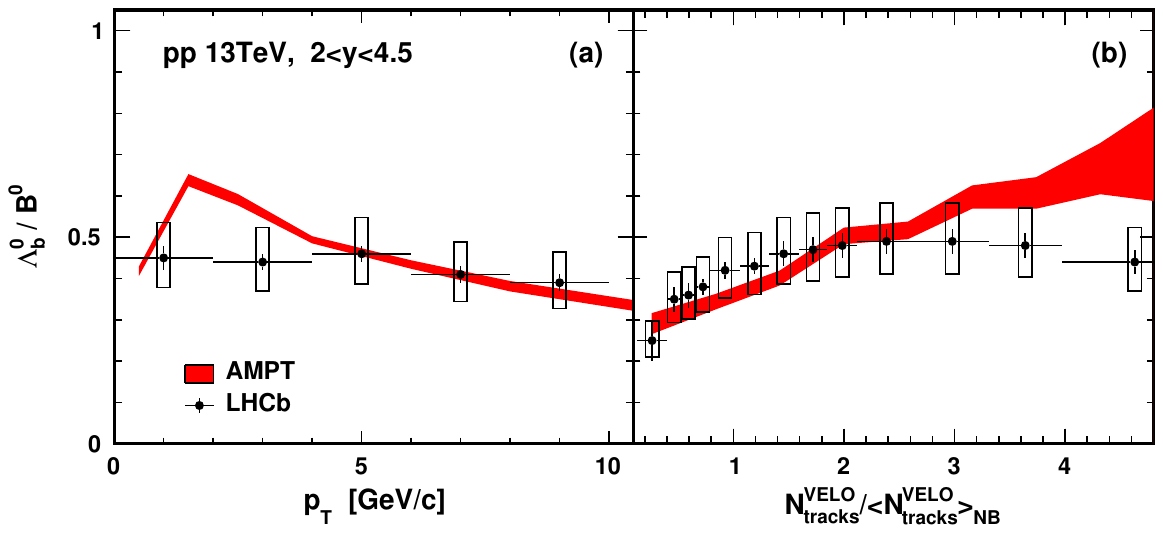}
		\caption{The $\Lambda_b^0/B^0$ ratios in $pp$ collisions at $\sqrt{s}=13$~TeV on forward-$y$ as a function of $p_T$ (a) and multiplicity (b). The red line and black points represent the AMPT and LHCb results~\cite{LHCb:2023wbo}, respectively. The vertical error bars on the LHCb points represent the quadratic sum of statistical and systematic uncertainties, and the boxes indicate global uncertainties associated with branching fractions.}
		\label{fig:LbB0}
	\end{center}
\end{figure*}

	\section{\label{sec:formalism}The AMPT Model}

    The AMPT model consists of four main components: initial conditions, partonic interactions, coalescence hadronization, and hadronic rescatterings. The initial conditions are generated using the PYTHIA8 Monte Carlo event generator~\cite{Bierlich:2022pfr} within the string-melting framework~\cite{Lin:2001zk,Lin:2002gc,Lin:2003iq}, incorporating sub-nucleon spatial fluctuations~\cite{Zheng:2021jrr}. The subsequent evolution of the resultant partonic matter produced from the PYTHIA8 initial conditions is simulated with Zhang’s Parton Cascade (ZPC) model~\cite{Zhang:1997ej}, which describes the microscopic two body parton scatterings. A parton rescattering cross section $\sigma=0.15\mathrm{mb}$ is employed, which gives a satisfactory description to the elliptic flow measurements in pp collisions~\cite{Zheng:2024xyv}. Hadronization is performed using a naive spatial coalescence scheme, in which quarks combine with their nearest spatial partners to form hadrons~\cite{He:2017tla}. An overall coalescence parameter $r_{BM}$ has been introduced to determine the relative probability for a quark to become a meson or a baryon in this model. The hadrons after coalescence may undergo further hadronic rescatterings implemented in the extended relativistic transport model (ART)~\cite{Li:1995pra,Li:2001xh}. 
    In this study, to address this issue, we have made improvements to both the initial conditions and the hadronization process, as detailed below:


It is found that the default parameters within PYTHIA8 produce a $b\bar{b}$ cross section that is higher than what has been measured experimentally at the top LHC energy as shown in Fig.~\ref{fig:bbbar}.  Figure.~\ref{fig:bbbar}(a) and (b) shows the $b\bar{b}$ cross section per rapidity ($y$) and pseudorapidity ($\eta$) in comparison with experimental measurements from ALICE~\cite{ALICE:2024xln} and LHCb~\cite{LHCb:2016qpe}, respectively. We observe that the default PYTHIA8 calculations with Monash tune indicated by the blue line significantly overestimate $b\bar{b}$ production. This discrepancy highlights that PYTHIA8's leading order framework does not inherently capture the full pQCD scale uncertainties and higher-order contributions that FONLL and GM-VFNS handle systematically. 
To obtain a realistic baseline for beauty production in the AMPT framework, we adjust the effective $b$ quark mass $m_b$ used in the quark production step, as has been done in Ref.~\cite{Bierlich:2023okq,Butenschoen:2016lpz,Nason:1999ta}. Through iterative adjustments of $m_{b}$ and comparisons with experimental data, we determined an optimal value of $m_{b}=6.6$ GeV$/c^2$. As shown in Fig.~\ref{fig:bbbar}, the results show substantial improvement over the default setting with $m_{b}=4.8$ GeV$/c^2$ and align more closely with experimental data when $m_{b}$ is increased to 6.6 GeV$/c^2$. We emphasize that the variation of $m_b$ is applied exclusively during the initial production stage implemented in PYTHIA8. After production, the mass is reset to the default value for all subsequent evolution processes, including partonic transport and hadronization. To illustrate the impact of this mass variation and place it in a broader theoretical context, we confront the resulting spectra with the widely established FONLL calculation in Appendix~\ref{sec:appendix2}. FONLL calculations incorporate higher order perturbative corrections and resummation effects that are not fully captured in leading order event generators. We find that increasing $m_b$ in PYTHIA8 effectively suppresses the beauty yield, which effectively mimics the higher order suppression effects characteristic of the FONLL predictions. This comparison demonstrates that the adopted tuning provides a phenomenologically controlled way to benchmark PYTHIA8 based initial conditions against the state of art perturbative QCD calculations.

In the AMPT coalescence hadronization model, the decision to form a meson or a baryon is based on the relative spatial separations between constituent partons in the partonic center-of-mass frame. For a given quark, the model identifies the closest antiquark, with \( d_{M} \) denoting the relative distance in the quark-antiquark rest frame. Simultaneously, it finds the two closest quarks, with \( d_{B} \) defined as the average of the three pairwise relative distances in the three-quark rest frame. A baryon is formed if the ratio $d_{B}/d_{M}$ is less than the coalescence model parameter $r_{BM}$~\cite{He:2017tla}. A larger $r_{BM}$ value corresponds to a higher baryon production yield. In the present work, we set $r_{BM}^{b}=1.2$ for beauty quarks, a value tuned to reproduce the measured $\Lambda_{b}$ production reported by the LHCb collaboration. The detailed comparison with experimental data, along with the systematic study underlying this choice, will be presented later in this paper. We note that $r_{BM}^{b}=1.2$ is slightly smaller than the value $r_{BM}^{c}=1.4$ used for charm quarks in our previous work, where the latter was found to successfully describe $\Lambda_{c}^{+}$ production~\cite{Zhang:2025pqu}. Compared to the light flavor coalescence parameters~\cite{Zhang:2025pqu,He:2017tla,Lin:2021mdn,Zhang:2019utb}, the large charm and beauty coalescence parameter is qualitatively consistent with expectations from the statistical hadronization model incorporating an augmented set of heavy flavor baryon states~\cite{He:2019tik}. 

It should be emphasized that the string melting version of AMPT integrates fragmentation and coalescence effects into a unified and internally consistent framework. In this approach, excited strings are first converted into primordial hadrons via string fragmentation, which serves as an intermediate step for implementing the string melting mechanism by decomposing these hadrons into their valence partons. The resulting partons subsequently evolve through the ZPC transport stage before hadronizing via a naive spatial coalescence algorithm that depends solely on spatial proximity. In low density systems, where partonic scatterings are rare and spatial correlations among partons originating from the same string are largely preserved, the coalescence procedure naturally recombines these partons back into their original primordial hadron configurations. As a result, fragmentation hadronization features are effectively recovered, even though the process is technically realized within the coalescence formalism. In this sense, fragmentation physics is effectively embedded in the modeling of the string melting process itself, and the subsequent parton transport stage provides a smooth transition from recombination dominated hadronization in dense environments to fragmentation like behavior in dilute systems within a single dynamical framework.

\begin{figure}[thbp]
	\begin{center}
		\includegraphics[width=0.48\textwidth]{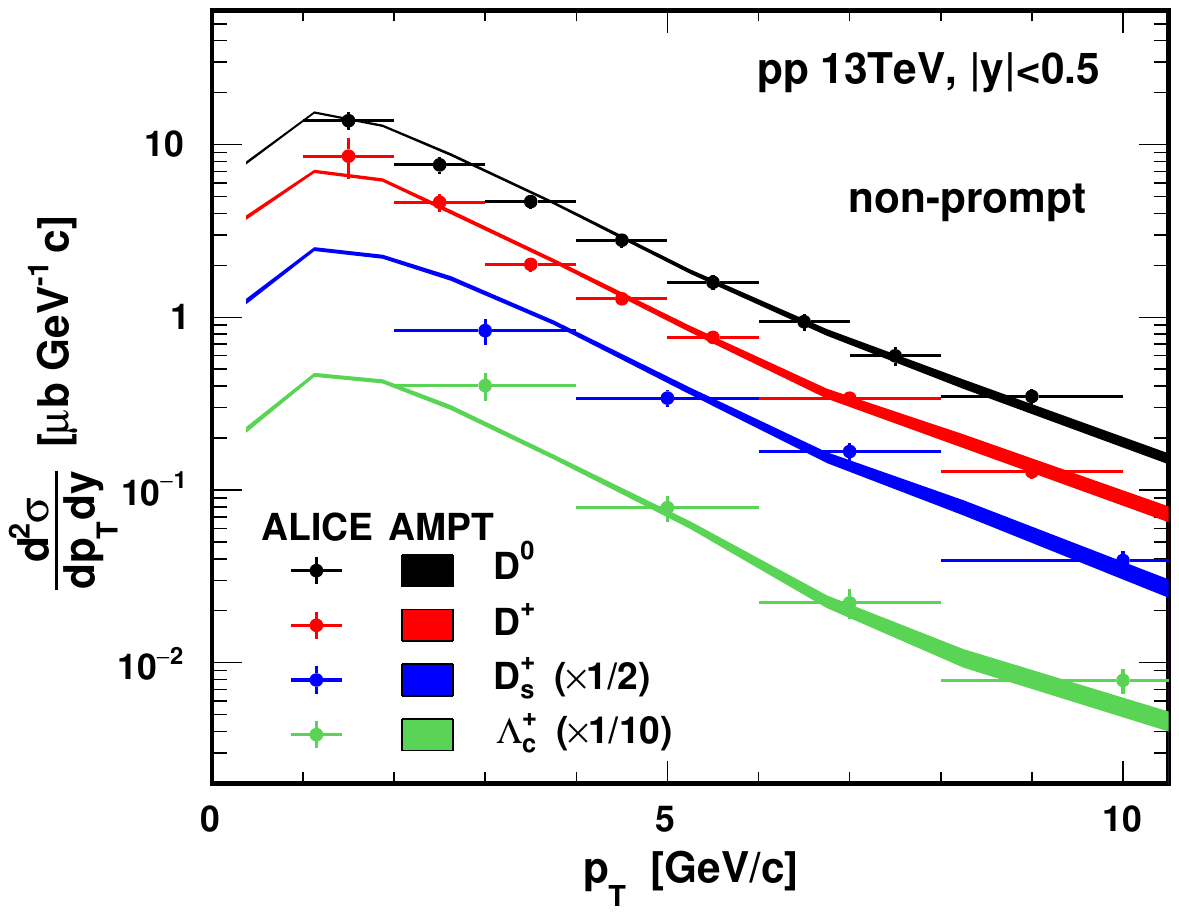}
		\caption{Double-differential cross section of non-prompt $c$-hadrons in $pp$ collisions at $\sqrt{s}=13$~TeV at mid-$y$, as a function of $p_T$. The $D^0$, $D^+$, $D_s^+$, and $\Lambda_c^+$ are shown in black, red, blue, and green, respectively. The $D_s^+$ and $\Lambda_c^+$ results were scaled by factors of 1/2 and 1/10 to allow better visual comparison. Lines represent AMPT results, while data points represent ALICE results~\cite{ALICE:2024xln, ALICE:2023wbx}.}
		\label{fig:nonprompt}
	\end{center}
\end{figure}
    
    \section{\label{sec:results}Results}
	\subsection{Beauty hadron produtions}\label{subsec:beauty}
    With the AMPT framework incorporating PYTHIA8 initial conditions, the tuned beauty-quark mass, and the optimized coalescence parameter $r_{BM}^b$, the model is now equipped with a consistent set of parameters that provide a realistic description of beauty hadron production. Building on this foundation, we proceed to investigate beauty hadrons and their non-prompt charm decay products in $pp$ collisions at $\sqrt{s}=$13 TeV. All the hadron cross sections presented in this work are calculated by averaging over particles and their corresponding antiparticles following the experimental analysis procedure outlined in Ref.~\cite{LHCb:2017vec,ALICE:2024xln,LHCb:2023wbo}.

	The $p_{T}$-differential cross sections of beauty hadrons in $pp$ collisions at $\sqrt{s} = 13$ TeV are presented in Fig.\ref{fig:bhadrons}. Panels (a) and (b) show the dependence on transverse momentum at mid- and forward rapidity, respectively, while panel (c) illustrates the rapidity dependence in the range $0 < p_T < 40$~GeV/$c$. The AMPT model results for $B^0$, $B^+$, $B_s$, and $\Lambda_b^0$ are shown as black, red, blue, and green lines, respectively, with solid and open black markers denoting $B^+$ measurements from LHCb~\cite{LHCb:2017vec} and ALICE~\cite{ALICE:2024xln}. The ALICE data are derived by scaling the $b\bar{b}$ cross section with a fragmentation fraction of 0.399~\cite{DELPHI:2011aa}. For visual clarity, the $B^0$ results are scaled by a factor of 3/4 in the figure. In Fig.~\ref{fig:bhadrons}(a) and (b), the $p_T$ distributions exhibit a relatively flat profile, consistent with beauty quarks produced predominantly via hard scattering processes involving large transverse momentum transfer. At forward rapidity, the AMPT predictions reproduce the measured $B^+$ spectra above $p_T \approx 2$GeV$/c$, while a noticeable excess appears at low $p_T$. This low-$p_T$ excess is attributed to threshold effects in the production of beauty quarks, amplified by the increased beauty quark mass. The large $m_b$ reduces the available phase space for beauty quark production, compressing the spectrum and shifting yield toward the lowest $p_T$ region. The effect is more pronounced at forward rapidity, where the typical partonic center-of-mass energies are closer to threshold, but it is largely absent in the midrapidity spectra. The calculated $B^0$ and $B^+$ spectra are nearly identical, as expected from isospin symmetry.  In Fig.~\ref{fig:bhadrons}(c), the rapidity distribution of $B^+$ from AMPT is consistent with LHCb data, providing confidence that the tuned model captures the main features of beauty hadron production at forward rapidity.
    

    
    In $pp$ collisions at the LHC energy~\cite{ALICE:2020wfu,ALICE:2023wbx}, the $\Lambda_{c}^{+}/D^{0}$ production ratio has been found to be significantly enhanced relative to measurements in $e^+e^-$ collisions~\cite{HFLAV:2019otj}, particularly at low and intermediate transverse momentum. Models incorporating a coalescence hadronization mechanism have been shown to describe this enhancement~\cite{ALICE:2020wfu}. Motivated by these findings, it is of interest to investigate whether analogous effects appear in the beauty sector.  Figure~\ref{fig:LbB0} shows the $\Lambda_b^0/B^0$ ratios in $pp$ collisions at $\sqrt{s}=$ 13 TeV at forward-$y$ as a function of $p_T$ (left) and relative charged particle multiplicity (right). The AMPT calculations are shown by the red line, while the black points correspond to measurements from the LHCb Collaboration~\cite{LHCb:2023wbo}.
    
    In Fig.\ref{fig:LbB0}(a), the AMPT calculations show a $\Lambda_b^0/B^0$ ratio varying with $p_T$ consistent with the value observed in the LHCb data and broadly compatible within experimental uncertainties. It is found that a mild excess appears around $p_T\approx$ 2 GeV/$c$ in the beauty baryon to meson ratio. This enhancement can be interpreted in terms of the kinematic consequences of the coalescence mechanism. The transverse momentum of a $\Lambda_b^0$ baryon arises from the vector sum of its three constituent quarks, in contrast to two for a $B^0$ meson, leading to a shifted baryon $p_T$ spectra and enhanced baryon to meson ratio at low $p_T$. The presence of this feature suggests a sizable contribution of coalescence effects to the beauty quark hadronization at low $p_T$. A qualitatively similar behavior has also been reported in hydrodynamics based simulations within the EPOS4HQ framework\cite{Zhao:2023ucp}.

    In Fig.~\ref{fig:LbB0}(b), the AMPT calculations reproduce the rising trend of the \(\Lambda_b^0/B^0\) ratio with charged-particle multiplicity observed by LHCb especially from low to intermediate multiplicity region. The horizontal axis represents the event activity, quantified as the number of reconstructed tracks normalized to the mean track multiplicity in the backward pseudorapidity interval \(2 < \eta < 5\). The enhancement of baryon production in AMPT increases with multiplicity, consistent with the data. At low multiplicity, the limited parton density favors hadronization through string fragmentation, yielding a relatively small baryon-to-meson ratio. With increasing multiplicity, the relative contribution of coalescence becomes more significant, leading to an enhanced ratio. The experimental measurements also show a tendency toward saturation of the ratio at high multiplicity, whereas this feature is less evident in the model. This saturation behavior aligns with expectations from statistical hadronization models~\cite{He:2019tik,Chen:2020drg,Dai:2024vjy}, where the system transitions from a canonical ensemble in low multiplicity $pp$ events to a grand canonical ensemble in larger system. The weaker saturation trend in AMPT implies that beauty quarks are less thermalized compared to data indications, with coalescence contributions continuing to grow with multiplicity. Nonetheless, the tuned AMPT model reasonably describes the multiplicity-dependent \(\Lambda_b^0/B^0\) ratio within experimental uncertainties, providing a robust basis for analyzing non-prompt charm hadron distributions.

\begin{figure*}[htbp]
	\begin{center}
		\includegraphics[width=1\textwidth]{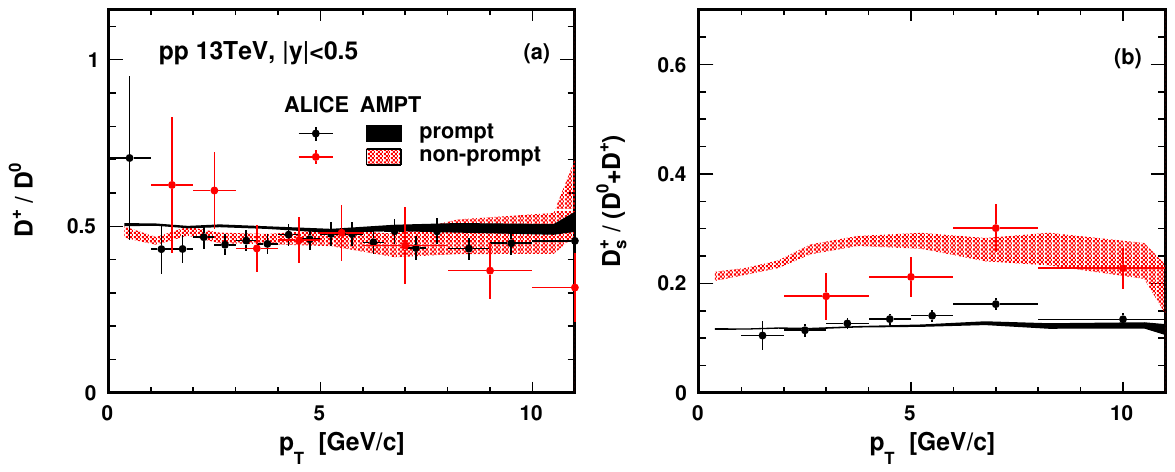}
		\caption{Ratios of $D^+/D^0$ (left) and $D_s^+/(D^0+D^+)$ (right) in $pp$ collisions at $\sqrt{s}=13$~TeV within $|y|<0.5$ as a function of $p_T$. Black and red represent prompt and non-prompt particles, respectively. The lines represent AMPT results, while the points correspond to ALICE results~\cite{ALICE:2024xln, ALICE:2023sgl}.}
		\label{fig:DmesonRatios}
	\end{center}
\end{figure*}

\begin{figure*}[htbp]
	\begin{center}
		\includegraphics[width=0.88\textwidth]{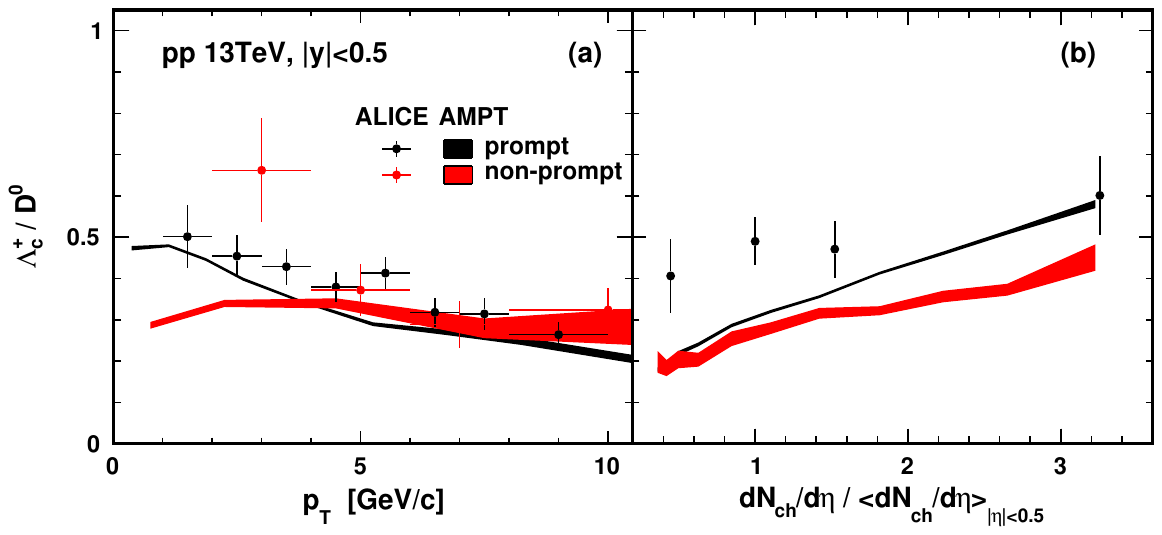}
		\caption{$\Lambda_{c}^{+}/D^{0}$ in $pp$ collisions at $\sqrt{s}=13$~TeV within $|y|<0.5$ as a function of $p_{T}$ (left) and multiplicity (right). Black and red represent prompt and non-prompt particles, respectively. The lines represent AMPT results, while the points correspond to ALICE results~\cite{ALICE:2021npz, ALICE:2023brx}.}
		\label{fig:LcD0}
	\end{center}
\end{figure*}

\begin{figure*}[htbp]
	\begin{center}
		\includegraphics[width=0.9\textwidth]{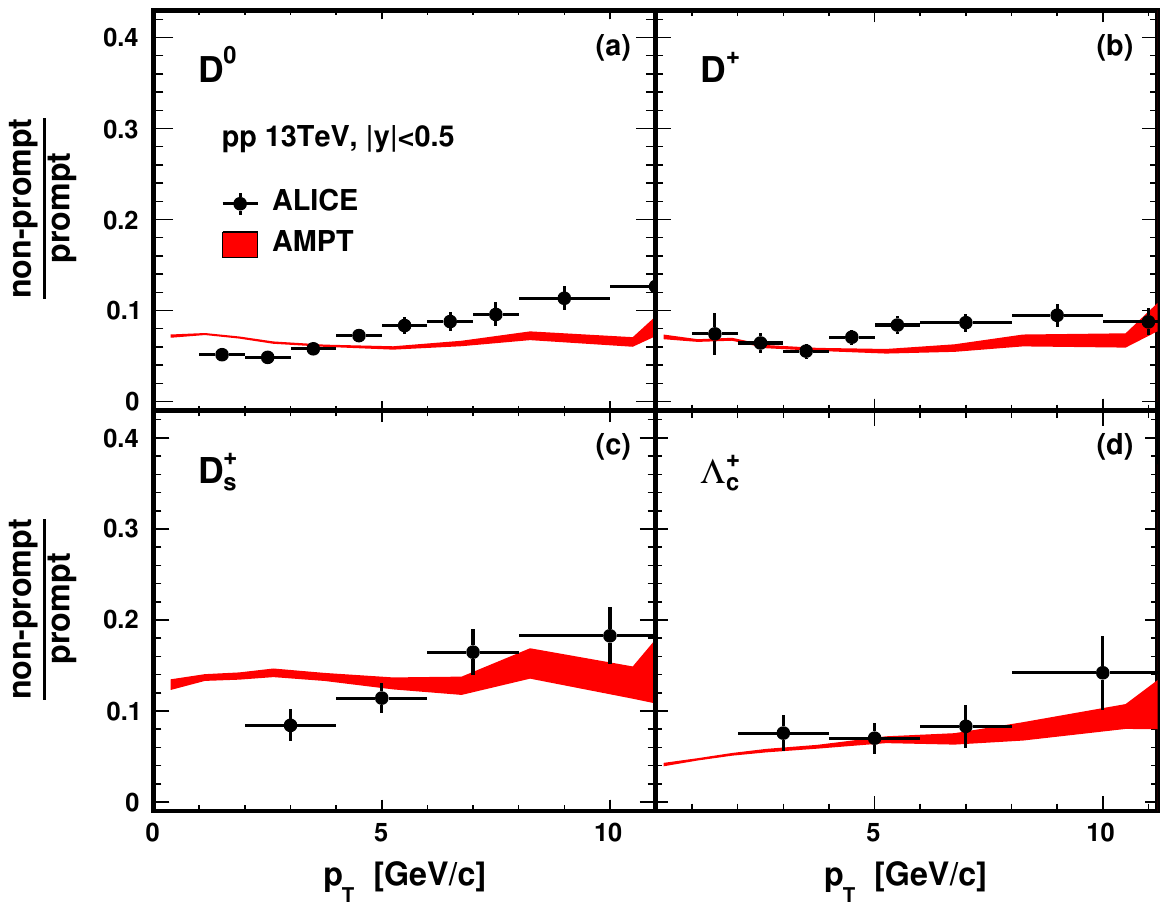}
		\caption{Non-prompt/prompt ratios in $pp$ collisions at $\sqrt{s}=13$~TeV within $|y|<0.5$ as a function of $p_T$. Panels (a)-(d) correspond to $D^0$, $D^+$, $D_s^+$, and $\Lambda_c^+$, respectively. Lines represent AMPT results, and points represent ALICE results~\cite{ALICE:2024xln, ALICE:2023wbx, ALICE:2021rzj}.}
		\label{fig:NPP_pt}
	\end{center}
\end{figure*}

\begin{figure*}[htbp]
	\begin{center}
		\includegraphics[width=0.9\textwidth]{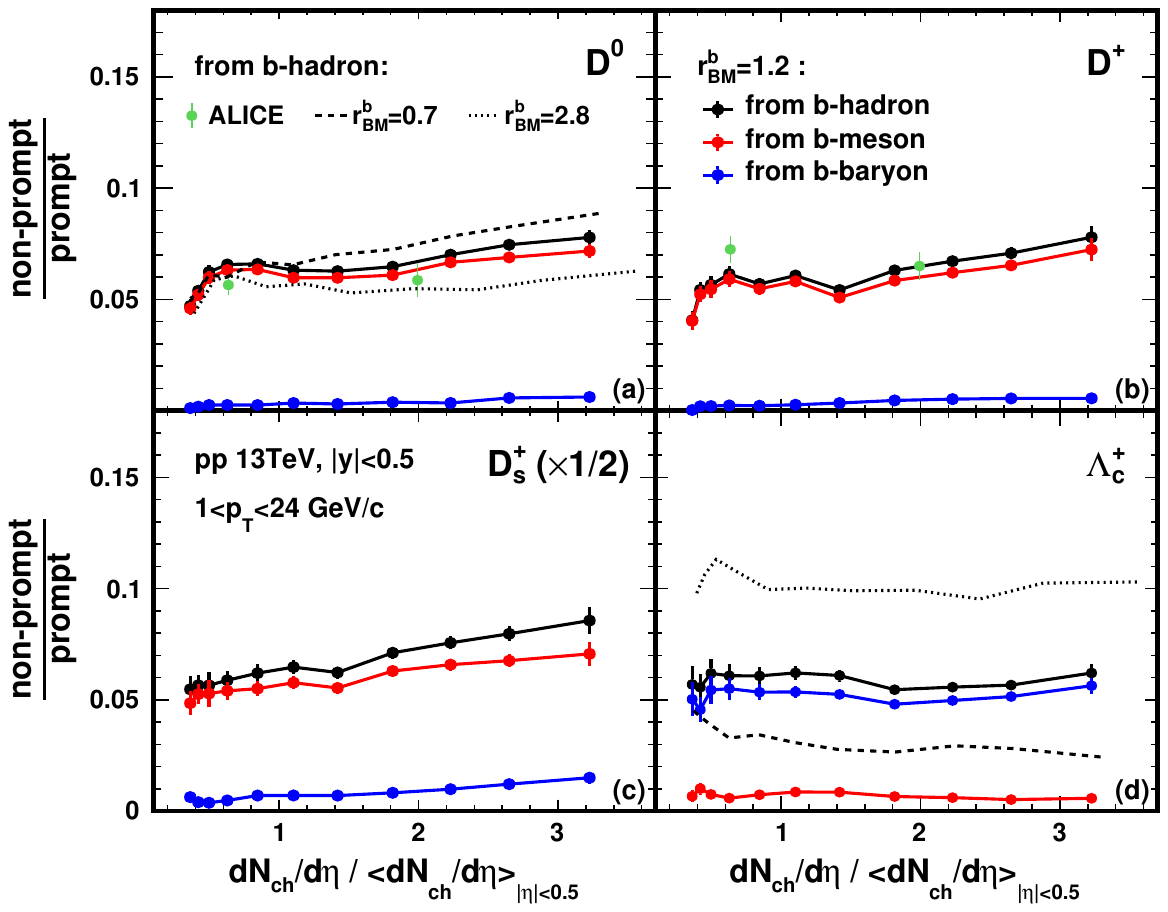}
		\caption{Non-prompt/prompt charm hadron ratios in $pp$ collisions at $\sqrt{s}=13$~TeV within $|y|<0.5$ for $1<p_T<24$~GeV$/c$ as a function of multiplicity. Panels (a)--(d) show results for $D^0$, $D^+$, $D_s^+$, and $\Lambda_c^+$, respectively. Solid black, red, and blue points denote non-prompt charm hadrons from all beauty hadron, only beauty meson, and only beauty baryon decays in AMPT of $r_{BM}^b=1.2$. Black dashed, dotted curves and green dots respectively represent non-prompt charm hadrons from all beauty hadron decays of $r_{BM}^b=0.7$, 2.8 and ALICE results~\cite{ALICE:2023brx}.}
		\label{fig:NPP_mult}
	\end{center}
\end{figure*}

	\subsection{Non-prompt charm hadron produtions}\label{subsec:nonprompt}

Since non-prompt charm hadrons originate from beauty-hadron decays, their spectra directly reflect the kinematic distributions and hadronization mechanisms of beauty quarks, making them a sensitive probe of beauty dynamics in small systems. After adjusting the input \(b\bar{b}\) cross section to align with experimental data, the AMPT framework yields an improved description of beauty-hadron production enabling reliable predictions for non-prompt charm hadrons. Figure~\ref{fig:nonprompt} shows the $p_T$ differential cross sections of non-prompt charm hadrons (\(D^0\), \(D^+\), \(D_s^+\), and \(\Lambda_c^+\)) in $pp$ collisions at \(\sqrt{s} = 13\) TeV in the mid-rapidity region (\(|y| < 0.5\)). The AMPT results, shown as black, red, blue, and green lines for \(D^0\), \(D^+\), \(D_s^+\), and \(\Lambda_c^+\), respectively, are compared to ALICE measurements~\cite{ALICE:2024xln,ALICE:2023wbx}. The calculations reproduce the experimental trends, with good agreement observed in particular for $D^0$, where the model predictions are consistent with the data within uncertainties across most of the measured $p_T$ region. A slight underestimation is seen for non-prompt \(\Lambda_c^+\). The comparison demonstrates that the tuned AMPT model provides a reasonable description of non-prompt charm-hadron production, establishing a baseline for further studies of charm-hadron ratios, baryon to meson production, and non-prompt to prompt fractions.

The relative yields of different charm meson species constrain the probabilities for a charm quark to combine with light or strange quarks, thereby testing isospin symmetry in the $D^+/D^0$ ratio and possible strangeness enhancement in the $D_s^+/(D^0+D^+)$ ratio. Separating prompt and non-prompt contributions further provides sensitivity to charm and beauty hadronization mechanisms. Figure~\ref{fig:DmesonRatios} shows the ratios of $D^+/D^0$ (left) and $D_s^+/(D^0+D^+)$ (right) as a function of $p_T$ in $pp$ collisions at $\sqrt{s}=13$ TeV at midrapidity. The AMPT model calculations are shown as lines, while the data points correspond to ALICE measurements~\cite{ALICE:2024xln,ALICE:2023sgl}. Prompt and non-prompt contributions are indicated in black and red, respectively.
    
In Fig.~\ref{fig:DmesonRatios}(a), the $D^+/D^0$ ratios from both AMPT and ALICE exhibit little dependence on $p_T$ for either prompt or non-prompt mesons, with values close to 0.5 across the measured range. The agreement between prompt and non-prompt results indicates that, within uncertainties, the hadronization of charm and beauty quarks with light $u$ and $d$ quarks proceeds with similar relative probabilities, consistent with expectations from isospin symmetry, and the decay kinematics of beauty hadrons do not alter this behavior. 
It is shown in Fig.~\ref{fig:DmesonRatios}(b) that the AMPT calculations for $D_s^+/(D^0+D^+)$ are compatible with ALICE data for both prompt and non-prompt charm hadrons. While the model reproduces the overall normalization, it predicts an essentially flat $p_T$ dependence, in contrast to the mild enhancement observed experimentally. The non-prompt strange to non-strange meson ratio lies significantly above the prompt one, which can be understood from the decay kinematics of beauty hadrons. Non-prompt $D_s^+$ mesons receive sizable contributions from both strange and non-strange $B$ meson decays, whereas non-prompt $D^0$ and $D^+$ mesons originate predominantly from non-strange $B$ mesons. 

To complement the study of beauty baryon-to-meson ratios, it is useful to investigate non-prompt charm hadrons, which act as effective proxies for beauty hadrons and allow direct comparison with their prompt counterparts. The relative size of prompt and non-prompt $\Lambda_{c}^{+}/D^{0}$ ratios provides sensitivity to the flavor dependence of hadronization. Figure~\ref{fig:LcD0} presents the  $\Lambda_{c}^{+}/D^{0}$ ratios in $pp$ collisions at $\sqrt{s}=13$ TeV as a function of transverse momentum (left) and charged-particle multiplicity (right). The AMPT results are shown as lines, while the ALICE measurements~\cite{ALICE:2021npz,ALICE:2021npz} are represented by points. Results for prompt and non-prompt charm hadrons are displayed in black and red, respectively.       

In Fig.~\ref{fig:LcD0}(a), the $p_{T}$ dependence of $\Lambda_{c}^{+}/D^{0}$ is presented. For the prompt component, both AMPT and ALICE data display an enhancement toward low $p_{T}$. Although the AMPT predictions lie slightly below the central values of the data, most results remain within experimental uncertainties, indicating that a choice of $r_{BM}^{c} = 1.4$ provides a reasonable description of the charm baryon to meson ratio. For the non-prompt component, the experimental data show a broadly similar trend to the prompt case, while AMPT reproduces the data well for $p_{T} > 4$ GeV/$c$ but falls below the data at low at lower $p_T$. The non-prompt $\Lambda_{c}^{+}/D^{0}$ in AMPT does not not exhibit an increase at low $p_T$. It is also noted that the decay kinematics of beauty hadrons smear the momentum distributions of the daughter charm hadrons, diluting the baryon enhancement at intermediate $p_T$ generated by coalescence shown in Fig.~\ref{fig:LbB0}(a). In particular, the AMPT prediction for non-prompt $\Lambda_{c}^{+}/D^{0}$ is systematically lower than the prompt ratio at small transverse momentum. This difference suggests distinct manifestations of coalescence dynamics in charm and beauty hadronization. 
Figure~\ref{fig:LcD0}(b) shows the multiplicity dependence of $\Lambda_{c}^{+}/D^{0}$ for prompt and non-prompt charm hadrons plotted as a function of the self-normalized charged-particle density $dN_{ch}/d\eta/\langle dN_{ch}/d\eta \rangle$. Both components exhibit a clear enhancement with increasing multiplicity, reflecting the growing role of partonic rescatterings and quark recombination in high-activity events. However, the prompt $\Lambda_{c}^{+}/D^{0}$ ratio rises more steeply than the non-prompt one, reflecting both the reduced sensitivity of beauty hadronization to event activity and the flavor dependence of the coalescence parameter in the AMPT framework.

    \subsection{Non-prompt to prompt charm hadron ratios}\label{subsec:nptop}
    
The ratio of non-prompt to prompt charm hadron ratios provide a key observable comparing the underlying kinematics of charm and beauty quarks. Figure~\ref{fig:NPP_pt} presents the non-prompt to prompt ratios in $pp$ collisions at $\sqrt{s}=13$ TeV at midrapidity as a function of $p_{T}$. Panels (a)-(d) correspond to $D^{0}$, $D^{+}$, $D_{s}^{+}$ and $\Lambda_c^{+}$, respectively. The curves represent AMPT results, while the data points denote ALICE measurements. The AMPT calculations reproduce the overall magnitude of the ratios reasonably well. This agreement arises because the large $m_b$ value reduces the yield of beauty hadrons, thereby lowering the non-prompt component. However, the model predictions for the non-prompt to prompt ratio show a relatively flat $p_T$ dependence, in contrast to the increasing trend observed in data. This discrepancy arises because the higher beauty-quark mass raises the production threshold, shifting a considerable fraction of beauty quarks to lower $p_T$. As a result, the shapes of the non-prompt and prompt charm-hadron spectra become more similar in the low $p_T$ region after the beauty hadron decay kinematics included, leading to a weaker rise of the non-prompt to prompt ratio with $p_T$ in AMPT. 
       
Beyond the transverse momentum dependence, the multiplicity dependence of the non-prompt over prompt charm hadron ratio provides an important probe of heavy flavor quark evolution in high density small system environments. Since non-prompt charm hadrons originate from both beauty meson and baryon decays, their relative contributions may vary with event activity. To disentangle these effects, the decay origins are tracked explicitly using the AMPT model. Figure~\ref{fig:NPP_mult} shows the non-prompt to prompt charm hadron ratios in $pp$ collisions at $\sqrt{s}=13$ TeV within $|y|<0.5$ for $1<p_T<24$~GeV$/c$ as a function of self-normalized event multiplicity, with panels (a)–(d) corresponding to $D^{0}$, $D^{+}$, $D_{s}^{+}$ and $\Lambda_c^{+}$, respectively. The non-prompt to prompt $D_s^+$ ratios were scaled by 1/2 for better visibility. Solid black, red, and blue points denote AMPT predictions for non-prompt charm hadrons from all beauty hadrons, beauty mesons only, and beauty baryons only, respectively. The solid black points are consistent with the sum of the latter two. Hollow black points indicate ALICE results for non-prompt charm hadrons from inclusive beauty hadron decays~\cite{ALICE:2023brx}. The dashed and dotted curves illustrate the effect of varying the beauty coalescence parameter to $r_{BM}^b=0.7$ and $2.8$, for the inclusive non-prompt to prompt ratios. For $D^{0}$ and $D^{+}$ from inclusive beauty hadron decay, AMPT reproduces the ALICE measurements well~\cite{ALICE:2023brx}. The AMPT calculations demonstrate that non-prompt charm mesons predominantly originate from beauty meson decays, while non-prompt charm baryons are largely produced in beauty baryon decays. This trend is consistent with the decay patterns discussed in the ALICE analysis~\cite{ALICE:2023brx}, highlighting the importance of the parent hadron composition in shaping the observed multiplicity dependence. The non-prompt $D_s^+$ receives a comparatively larger contribution from beauty baryon decays than the $D^0$ and $D^+$ mesons, although its overall multiplicity dependence remains similar to the other meson species. A contrast emerges between multiplicity dependences of meson and baryon species. AMPT predicts a modest increase of the non-prompt to prompt ratio with multiplicity for charm mesons, while for $\Lambda_c^{+}$ this behavior becomes less compelling. This discrepancy represents the sensitivity of the multiplicity dependence of the non-prompt to prompt charm hadron ratio to the relative values of the coalescence parameter $r_{BM}$ for charm and beauty. Varying the $r_{BM}^b$ parameter significantly alters the multiplicity dependence of non-prompt to prompt particle ratios as indicated by the dashed and dotted lines corresponding to $r_{BM}^b$ of 0.7 and 2.8. Smaller value $r_{BM}^b=0.7$ suppresses beauty baryon formation, leading to a stronger rise of the $D^0$ ratios with multiplicity but a marked decrease for $\Lambda_c^{+}$. Conversely, a larger value $r_{BM}^b=2.8$ enhances beauty baryon production, flattening the multiplicity trend for mesons while boosting the baryon channel.
A more systematic decomposition into primordial hadron production, coalescence probabilities, and decay kinematics is presented in Appendix~\ref{sec:appendix}. The tuned $r_{BM}^b$ for beauty hadrons naturally leads to a slower growth of the non-prompt to prompt ratio varying with multiplicity consistent with the observations found in experimental data. Future precision measurements of non-prompt baryon to meson ratios as a function of multiplicity could thus provide a powerful constraint on flavor dependent coalescence dynamics, helping to bridge microscopic transport approaches with statistical descriptions of heavy flavor hadronization.

    \section{\label{sec:Summary}Summary}
    
We present a comprehensive study of beauty hadron and non-prompt charm hadron production in $pp$ collisions at $\sqrt{s}=13$ TeV within the framework of the AMPT model coupled with PYTHIA8 initial conditions. To improve the description of beauty related observables within AMPT, two targeted modifications are introduced. First, the beauty quark mass is increased to $m_{b}=6.6$ GeV$/c^2$ in order to reproduce the measured $b\bar{b}$ production cross section, consistent with phenomenological adjustments often applied in FONLL and related QCD calculations~\cite{Cacciari:2012ny}. Second, a flavor dependent coalescence parameter is applied, with $r_{BM}^b=1.2$ for beauty quarks, guided by LHCb measurements of beauty baryon to meson ratios~\cite{LHCb:2017vec,LHCb:2023wbo}. With these refinements, AMPT provides an improved description of beauty hadron spectra and non-prompt charm hadron production at both forward and mid-rapidity, reproducing their transverse momentum and multiplicity dependences.

It is shown in this study that the multiplicity dependence of non-prompt to prompt charm hadron ratios can be sensitive to the beauty quark coalescence dynamics. A reduced beauty flavor coalescence parameter $r_{BM}^b$ suppresses the beauty baryon formation, thereby enhancing the beauty meson decay channels, which leads to a stronger increase with multiplicity for non-prompt to prompt charm meson ratios. Conversely, an enlarged $r_{BM}^b$ value promotes beauty baryon production, flattening the meson ratios and amplifying the baryon enhancement. These findings establish non-prompt charm hadrons as powerful observables for probing heavy-flavor hadronization in small systems. Their sensitivity to the interplay between quark production, coalescence dynamics, and decay kinematics makes them ideal for disentangling the differences between charm and beauty hadronization mechanisms. 

Future high precision measurements of non-prompt baryon and meson productions as a function of event activity will thus be essential for constraining theoretical models, bridging transport based approaches like AMPT with statistical hadronization descriptions, and advancing our understanding of heavy quark dynamics in the dense environment of high multiplicity $pp$ collisions. The AMPT framework, by providing a unified treatment of heavy flavor production and bulk medium evolution, offers a unique platform for investigating heavy quark collectivity phenomena and opens a path toward a deeper understanding of the role of beauty quarks in the evolution of small systems.

	\begin{acknowledgments}
This work was supported by the National Key Research and Development Program of China (Grant No. 2024YFA1610804), the National Natural Science Foundation of China (Nos. 12205259, 12147101, 12275103, 12061141008), the Fundamental Research Funds for the Central Universities, China University of Geosciences(Wuhan) with No. G1323523064 and the Innovation Fund of Key Laboratory of Quark and Lepton Physics QLPL2025P01.
		
	\end{acknowledgments}
	
	
	\bibliography{main}

\begin{thebibliography}{54}%
\makeatletter
\providecommand \@ifxundefined [1]{%
 \@ifx{#1\undefined}
}%
\providecommand \@ifnum [1]{%
 \ifnum #1\expandafter \@firstoftwo
 \else \expandafter \@secondoftwo
 \fi
}%
\providecommand \@ifx [1]{%
 \ifx #1\expandafter \@firstoftwo
 \else \expandafter \@secondoftwo
 \fi
}%
\providecommand \natexlab [1]{#1}%
\providecommand \enquote  [1]{``#1''}%
\providecommand \bibnamefont  [1]{#1}%
\providecommand \bibfnamefont [1]{#1}%
\providecommand \citenamefont [1]{#1}%
\providecommand \href@noop [0]{\@secondoftwo}%
\providecommand \href [0]{\begingroup \@sanitize@url \@href}%
\providecommand \@href[1]{\@@startlink{#1}\@@href}%
\providecommand \@@href[1]{\endgroup#1\@@endlink}%
\providecommand \@sanitize@url [0]{\catcode `\\12\catcode `\$12\catcode
  `\&12\catcode `\#12\catcode `\^12\catcode `\_12\catcode `\%12\relax}%
\providecommand \@@startlink[1]{}%
\providecommand \@@endlink[0]{}%
\providecommand \url  [0]{\begingroup\@sanitize@url \@url }%
\providecommand \@url [1]{\endgroup\@href {#1}{\urlprefix }}%
\providecommand \urlprefix  [0]{URL }%
\providecommand \Eprint [0]{\href }%
\providecommand \doibase [0]{http://dx.doi.org/}%
\providecommand \selectlanguage [0]{\@gobble}%
\providecommand \bibinfo  [0]{\@secondoftwo}%
\providecommand \bibfield  [0]{\@secondoftwo}%
\providecommand \translation [1]{[#1]}%
\providecommand \BibitemOpen [0]{}%
\providecommand \bibitemStop [0]{}%
\providecommand \bibitemNoStop [0]{.\EOS\space}%
\providecommand \EOS [0]{\spacefactor3000\relax}%
\providecommand \BibitemShut  [1]{\csname bibitem#1\endcsname}%
\let\auto@bib@innerbib\@empty
\bibitem [{\citenamefont {Andronic}\ \emph {et~al.}(2016)\citenamefont
  {Andronic} \emph {et~al.}}]{Andronic:2015wma}%
  \BibitemOpen
  \bibfield  {author} {\bibinfo {author} {\bibfnamefont {A.}~\bibnamefont
  {Andronic}} \emph {et~al.},\ }\href {\doibase 10.1140/epjc/s10052-015-3819-5}
  {\bibfield  {journal} {\bibinfo  {journal} {Eur. Phys. J. C}\ }\textbf
  {\bibinfo {volume} {76}},\ \bibinfo {pages} {107} (\bibinfo {year} {2016})},\
  \Eprint {http://arxiv.org/abs/1506.03981} {arXiv:1506.03981 [nucl-ex]}
  \BibitemShut {NoStop}%
\bibitem [{\citenamefont {Collins}\ \emph {et~al.}(1989)\citenamefont
  {Collins}, \citenamefont {Soper},\ and\ \citenamefont
  {Sterman}}]{Collins:1989gx}%
  \BibitemOpen
  \bibfield  {author} {\bibinfo {author} {\bibfnamefont {J.~C.}\ \bibnamefont
  {Collins}}, \bibinfo {author} {\bibfnamefont {D.~E.}\ \bibnamefont {Soper}},
  \ and\ \bibinfo {author} {\bibfnamefont {G.~F.}\ \bibnamefont {Sterman}},\
  }\href {\doibase 10.1142/9789814503266_0001} {\bibfield  {journal} {\bibinfo
  {journal} {Adv. Ser. Direct. High Energy Phys.}\ }\textbf {\bibinfo {volume}
  {5}},\ \bibinfo {pages} {1} (\bibinfo {year} {1989})},\ \Eprint
  {http://arxiv.org/abs/hep-ph/0409313} {arXiv:hep-ph/0409313} \BibitemShut
  {NoStop}%
\bibitem [{\citenamefont {Apolin{\'a}rio}\ \emph {et~al.}(2022)\citenamefont
  {Apolin{\'a}rio}, \citenamefont {Lee},\ and\ \citenamefont
  {Winn}}]{Apolinario:2022vzg}%
  \BibitemOpen
  \bibfield  {author} {\bibinfo {author} {\bibfnamefont {L.}~\bibnamefont
  {Apolin{\'a}rio}}, \bibinfo {author} {\bibfnamefont {Y.-J.}\ \bibnamefont
  {Lee}}, \ and\ \bibinfo {author} {\bibfnamefont {M.}~\bibnamefont {Winn}},\
  }\href {\doibase 10.1016/j.ppnp.2022.103990} {\bibfield  {journal} {\bibinfo
  {journal} {Prog. Part. Nucl. Phys.}\ }\textbf {\bibinfo {volume} {127}},\
  \bibinfo {pages} {103990} (\bibinfo {year} {2022})},\ \Eprint
  {http://arxiv.org/abs/2203.16352} {arXiv:2203.16352 [hep-ph]} \BibitemShut
  {NoStop}%
\bibitem [{\citenamefont {Bedjidian}\ \emph {et~al.}(2004)\citenamefont
  {Bedjidian} \emph {et~al.}}]{Bedjidian:2004gd}%
  \BibitemOpen
  \bibfield  {author} {\bibinfo {author} {\bibfnamefont {M.}~\bibnamefont
  {Bedjidian}} \emph {et~al.}\ }(\bibinfo {year} {2004})\ \Eprint
  {http://arxiv.org/abs/hep-ph/0311048} {arXiv:hep-ph/0311048} \BibitemShut
  {NoStop}%
\bibitem [{\citenamefont {Schweda}(2014)}]{Schweda:2014tya}%
  \BibitemOpen
  \bibfield  {author} {\bibinfo {author} {\bibfnamefont {K.~O.}\ \bibnamefont
  {Schweda}},\ }\emph {\bibinfo {title} {{Prompt production of D mesons with
  ALICE at the LHC}}},\ \href@noop {} {Ph.D. thesis},\ \bibinfo  {school}
  {Heidelberg U.} (\bibinfo {year} {2014}),\ \Eprint
  {http://arxiv.org/abs/1402.1370} {arXiv:1402.1370 [nucl-ex]} \BibitemShut
  {NoStop}%
\bibitem [{\citenamefont {Cacciari}\ \emph {et~al.}(2005)\citenamefont
  {Cacciari}, \citenamefont {Nason},\ and\ \citenamefont
  {Vogt}}]{Cacciari:2005rk}%
  \BibitemOpen
  \bibfield  {author} {\bibinfo {author} {\bibfnamefont {M.}~\bibnamefont
  {Cacciari}}, \bibinfo {author} {\bibfnamefont {P.}~\bibnamefont {Nason}}, \
  and\ \bibinfo {author} {\bibfnamefont {R.}~\bibnamefont {Vogt}},\ }\href
  {\doibase 10.1103/PhysRevLett.95.122001} {\bibfield  {journal} {\bibinfo
  {journal} {Phys. Rev. Lett.}\ }\textbf {\bibinfo {volume} {95}},\ \bibinfo
  {pages} {122001} (\bibinfo {year} {2005})},\ \Eprint
  {http://arxiv.org/abs/hep-ph/0502203} {arXiv:hep-ph/0502203} \BibitemShut
  {NoStop}%
\bibitem [{\citenamefont {Kniehl}\ \emph {et~al.}(2005)\citenamefont {Kniehl},
  \citenamefont {Kramer}, \citenamefont {Schienbein},\ and\ \citenamefont
  {Spiesberger}}]{Kniehl:2004fy}%
  \BibitemOpen
  \bibfield  {author} {\bibinfo {author} {\bibfnamefont {B.~A.}\ \bibnamefont
  {Kniehl}}, \bibinfo {author} {\bibfnamefont {G.}~\bibnamefont {Kramer}},
  \bibinfo {author} {\bibfnamefont {I.}~\bibnamefont {Schienbein}}, \ and\
  \bibinfo {author} {\bibfnamefont {H.}~\bibnamefont {Spiesberger}},\ }\href
  {\doibase 10.1103/PhysRevD.71.014018} {\bibfield  {journal} {\bibinfo
  {journal} {Phys. Rev. D}\ }\textbf {\bibinfo {volume} {71}},\ \bibinfo
  {pages} {014018} (\bibinfo {year} {2005})},\ \Eprint
  {http://arxiv.org/abs/hep-ph/0410289} {arXiv:hep-ph/0410289} \BibitemShut
  {NoStop}%
\bibitem [{\citenamefont {Helenius}\ and\ \citenamefont
  {Paukkunen}(2018)}]{Helenius:2018uul}%
  \BibitemOpen
  \bibfield  {author} {\bibinfo {author} {\bibfnamefont {I.}~\bibnamefont
  {Helenius}}\ and\ \bibinfo {author} {\bibfnamefont {H.}~\bibnamefont
  {Paukkunen}},\ }\href {\doibase 10.1007/JHEP05(2018)196} {\bibfield
  {journal} {\bibinfo  {journal} {JHEP}\ }\textbf {\bibinfo {volume} {05}},\
  \bibinfo {pages} {196} (\bibinfo {year} {2018})},\ \Eprint
  {http://arxiv.org/abs/1804.03557} {arXiv:1804.03557 [hep-ph]} \BibitemShut
  {NoStop}%
\bibitem [{\citenamefont {Mangano}\ \emph {et~al.}(1992)\citenamefont
  {Mangano}, \citenamefont {Nason},\ and\ \citenamefont
  {Ridolfi}}]{Mangano:1991jk}%
  \BibitemOpen
  \bibfield  {author} {\bibinfo {author} {\bibfnamefont {M.~L.}\ \bibnamefont
  {Mangano}}, \bibinfo {author} {\bibfnamefont {P.}~\bibnamefont {Nason}}, \
  and\ \bibinfo {author} {\bibfnamefont {G.}~\bibnamefont {Ridolfi}},\ }\href
  {\doibase 10.1016/0550-3213(92)90435-E} {\bibfield  {journal} {\bibinfo
  {journal} {Nucl. Phys. B}\ }\textbf {\bibinfo {volume} {373}},\ \bibinfo
  {pages} {295} (\bibinfo {year} {1992})}\BibitemShut {NoStop}%
\bibitem [{\citenamefont {Acharya}\ \emph
  {et~al.}(2023{\natexlab{a}})\citenamefont {Acharya} \emph
  {et~al.}}]{ALICE:2023sgl}%
  \BibitemOpen
  \bibfield  {author} {\bibinfo {author} {\bibfnamefont {S.}~\bibnamefont
  {Acharya}} \emph {et~al.} (\bibinfo {collaboration} {ALICE}),\ }\href
  {\doibase 10.1007/JHEP12(2023)086} {\bibfield  {journal} {\bibinfo  {journal}
  {JHEP}\ }\textbf {\bibinfo {volume} {12}},\ \bibinfo {pages} {086} (\bibinfo
  {year} {2023}{\natexlab{a}})},\ \Eprint {http://arxiv.org/abs/2308.04877}
  {arXiv:2308.04877 [hep-ex]} \BibitemShut {NoStop}%
\bibitem [{\citenamefont {Abualrob}\ \emph {et~al.}(2025)\citenamefont
  {Abualrob} \emph {et~al.}}]{ALICE:2025wrq}%
  \BibitemOpen
  \bibfield  {author} {\bibinfo {author} {\bibfnamefont {I.~J.}\ \bibnamefont
  {Abualrob}} \emph {et~al.} (\bibinfo {collaboration} {ALICE}),\ }\href@noop
  {} {\  (\bibinfo {year} {2025})},\ \Eprint {http://arxiv.org/abs/2508.09955}
  {arXiv:2508.09955 [nucl-ex]} \BibitemShut {NoStop}%
\bibitem [{\citenamefont {Altmann}\ \emph {et~al.}(2025)\citenamefont
  {Altmann}, \citenamefont {Dubla}, \citenamefont {Greco}, \citenamefont
  {Rossi},\ and\ \citenamefont {Skands}}]{Altmann:2024kwx}%
  \BibitemOpen
  \bibfield  {author} {\bibinfo {author} {\bibfnamefont {J.}~\bibnamefont
  {Altmann}}, \bibinfo {author} {\bibfnamefont {A.}~\bibnamefont {Dubla}},
  \bibinfo {author} {\bibfnamefont {V.}~\bibnamefont {Greco}}, \bibinfo
  {author} {\bibfnamefont {A.}~\bibnamefont {Rossi}}, \ and\ \bibinfo {author}
  {\bibfnamefont {P.}~\bibnamefont {Skands}},\ }\href {\doibase
  10.1140/epjc/s10052-024-13641-5} {\bibfield  {journal} {\bibinfo  {journal}
  {Eur. Phys. J. C}\ }\textbf {\bibinfo {volume} {85}},\ \bibinfo {pages} {16}
  (\bibinfo {year} {2025})},\ \Eprint {http://arxiv.org/abs/2405.19137}
  {arXiv:2405.19137 [hep-ph]} \BibitemShut {NoStop}%
\bibitem [{\citenamefont {Chen}\ and\ \citenamefont {He}(2021)}]{Chen:2020drg}%
  \BibitemOpen
  \bibfield  {author} {\bibinfo {author} {\bibfnamefont {Y.}~\bibnamefont
  {Chen}}\ and\ \bibinfo {author} {\bibfnamefont {M.}~\bibnamefont {He}},\
  }\href {\doibase 10.1016/j.physletb.2021.136144} {\bibfield  {journal}
  {\bibinfo  {journal} {Phys. Lett. B}\ }\textbf {\bibinfo {volume} {815}},\
  \bibinfo {pages} {136144} (\bibinfo {year} {2021})},\ \Eprint
  {http://arxiv.org/abs/2011.14328} {arXiv:2011.14328 [hep-ph]} \BibitemShut
  {NoStop}%
\bibitem [{\citenamefont {Dai}\ \emph {et~al.}(2024)\citenamefont {Dai},
  \citenamefont {Zhao},\ and\ \citenamefont {He}}]{Dai:2024vjy}%
  \BibitemOpen
  \bibfield  {author} {\bibinfo {author} {\bibfnamefont {Y.}~\bibnamefont
  {Dai}}, \bibinfo {author} {\bibfnamefont {S.}~\bibnamefont {Zhao}}, \ and\
  \bibinfo {author} {\bibfnamefont {M.}~\bibnamefont {He}},\ }\href {\doibase
  10.1103/PhysRevC.110.034905} {\bibfield  {journal} {\bibinfo  {journal}
  {Phys. Rev. C}\ }\textbf {\bibinfo {volume} {110}},\ \bibinfo {pages}
  {034905} (\bibinfo {year} {2024})},\ \Eprint
  {http://arxiv.org/abs/2402.03692} {arXiv:2402.03692 [hep-ph]} \BibitemShut
  {NoStop}%
\bibitem [{\citenamefont {He}\ and\ \citenamefont {Rapp}(2020)}]{He:2019vgs}%
  \BibitemOpen
  \bibfield  {author} {\bibinfo {author} {\bibfnamefont {M.}~\bibnamefont
  {He}}\ and\ \bibinfo {author} {\bibfnamefont {R.}~\bibnamefont {Rapp}},\
  }\href {\doibase 10.1103/PhysRevLett.124.042301} {\bibfield  {journal}
  {\bibinfo  {journal} {Phys. Rev. Lett.}\ }\textbf {\bibinfo {volume} {124}},\
  \bibinfo {pages} {042301} (\bibinfo {year} {2020})},\ \Eprint
  {http://arxiv.org/abs/1905.09216} {arXiv:1905.09216 [nucl-th]} \BibitemShut
  {NoStop}%
\bibitem [{\citenamefont {He}\ and\ \citenamefont {Rapp}(2019)}]{He:2019tik}%
  \BibitemOpen
  \bibfield  {author} {\bibinfo {author} {\bibfnamefont {M.}~\bibnamefont
  {He}}\ and\ \bibinfo {author} {\bibfnamefont {R.}~\bibnamefont {Rapp}},\
  }\href {\doibase 10.1016/j.physletb.2019.06.004} {\bibfield  {journal}
  {\bibinfo  {journal} {Phys. Lett. B}\ }\textbf {\bibinfo {volume} {795}},\
  \bibinfo {pages} {117} (\bibinfo {year} {2019})},\ \Eprint
  {http://arxiv.org/abs/1902.08889} {arXiv:1902.08889 [nucl-th]} \BibitemShut
  {NoStop}%
\bibitem [{\citenamefont {Zhao}\ \emph
  {et~al.}(2024{\natexlab{a}})\citenamefont {Zhao} \emph
  {et~al.}}]{Zhao:2023nrz}%
  \BibitemOpen
  \bibfield  {author} {\bibinfo {author} {\bibfnamefont {J.}~\bibnamefont
  {Zhao}} \emph {et~al.},\ }\href {\doibase 10.1103/PhysRevC.109.054912}
  {\bibfield  {journal} {\bibinfo  {journal} {Phys. Rev. C}\ }\textbf {\bibinfo
  {volume} {109}},\ \bibinfo {pages} {054912} (\bibinfo {year}
  {2024}{\natexlab{a}})},\ \Eprint {http://arxiv.org/abs/2311.10621}
  {arXiv:2311.10621 [hep-ph]} \BibitemShut {NoStop}%
\bibitem [{\citenamefont {Minissale}\ \emph {et~al.}(2021)\citenamefont
  {Minissale}, \citenamefont {Plumari},\ and\ \citenamefont
  {Greco}}]{Minissale:2020bif}%
  \BibitemOpen
  \bibfield  {author} {\bibinfo {author} {\bibfnamefont {V.}~\bibnamefont
  {Minissale}}, \bibinfo {author} {\bibfnamefont {S.}~\bibnamefont {Plumari}},
  \ and\ \bibinfo {author} {\bibfnamefont {V.}~\bibnamefont {Greco}},\ }\href
  {\doibase 10.1016/j.physletb.2021.136622} {\bibfield  {journal} {\bibinfo
  {journal} {Phys. Lett. B}\ }\textbf {\bibinfo {volume} {821}},\ \bibinfo
  {pages} {136622} (\bibinfo {year} {2021})},\ \Eprint
  {http://arxiv.org/abs/2012.12001} {arXiv:2012.12001 [hep-ph]} \BibitemShut
  {NoStop}%
\bibitem [{\citenamefont {Minissale}\ \emph {et~al.}(2025)\citenamefont
  {Minissale}, \citenamefont {Greco},\ and\ \citenamefont
  {Plumari}}]{Minissale:2024gxx}%
  \BibitemOpen
  \bibfield  {author} {\bibinfo {author} {\bibfnamefont {V.}~\bibnamefont
  {Minissale}}, \bibinfo {author} {\bibfnamefont {V.}~\bibnamefont {Greco}}, \
  and\ \bibinfo {author} {\bibfnamefont {S.}~\bibnamefont {Plumari}},\ }\href
  {\doibase 10.1016/j.physletb.2024.139190} {\bibfield  {journal} {\bibinfo
  {journal} {Phys. Lett. B}\ }\textbf {\bibinfo {volume} {860}},\ \bibinfo
  {pages} {139190} (\bibinfo {year} {2025})},\ \Eprint
  {http://arxiv.org/abs/2405.19244} {arXiv:2405.19244 [hep-ph]} \BibitemShut
  {NoStop}%
\bibitem [{\citenamefont {Zhao}\ \emph
  {et~al.}(2024{\natexlab{b}})\citenamefont {Zhao}, \citenamefont {Aichelin},
  \citenamefont {Gossiaux}, \citenamefont {Ozvenchuk},\ and\ \citenamefont
  {Werner}}]{Zhao:2024ecc}%
  \BibitemOpen
  \bibfield  {author} {\bibinfo {author} {\bibfnamefont {J.}~\bibnamefont
  {Zhao}}, \bibinfo {author} {\bibfnamefont {J.}~\bibnamefont {Aichelin}},
  \bibinfo {author} {\bibfnamefont {P.~B.}\ \bibnamefont {Gossiaux}}, \bibinfo
  {author} {\bibfnamefont {V.}~\bibnamefont {Ozvenchuk}}, \ and\ \bibinfo
  {author} {\bibfnamefont {K.}~\bibnamefont {Werner}},\ }\href {\doibase
  10.1103/PhysRevC.110.024909} {\bibfield  {journal} {\bibinfo  {journal}
  {Phys. Rev. C}\ }\textbf {\bibinfo {volume} {110}},\ \bibinfo {pages}
  {024909} (\bibinfo {year} {2024}{\natexlab{b}})},\ \Eprint
  {http://arxiv.org/abs/2401.17096} {arXiv:2401.17096 [hep-ph]} \BibitemShut
  {NoStop}%
\bibitem [{\citenamefont {Bierlich}\ \emph {et~al.}(2015)\citenamefont
  {Bierlich}, \citenamefont {Gustafson}, \citenamefont {L{\"o}nnblad},\ and\
  \citenamefont {Tarasov}}]{Bierlich:2014xba}%
  \BibitemOpen
  \bibfield  {author} {\bibinfo {author} {\bibfnamefont {C.}~\bibnamefont
  {Bierlich}}, \bibinfo {author} {\bibfnamefont {G.}~\bibnamefont {Gustafson}},
  \bibinfo {author} {\bibfnamefont {L.}~\bibnamefont {L{\"o}nnblad}}, \ and\
  \bibinfo {author} {\bibfnamefont {A.}~\bibnamefont {Tarasov}},\ }\href
  {\doibase 10.1007/JHEP03(2015)148} {\bibfield  {journal} {\bibinfo  {journal}
  {JHEP}\ }\textbf {\bibinfo {volume} {03}},\ \bibinfo {pages} {148} (\bibinfo
  {year} {2015})},\ \Eprint {http://arxiv.org/abs/1412.6259} {arXiv:1412.6259
  [hep-ph]} \BibitemShut {NoStop}%
\bibitem [{\citenamefont {Bierlich}\ and\ \citenamefont
  {Christiansen}(2015)}]{Bierlich:2015rha}%
  \BibitemOpen
  \bibfield  {author} {\bibinfo {author} {\bibfnamefont {C.}~\bibnamefont
  {Bierlich}}\ and\ \bibinfo {author} {\bibfnamefont {J.~R.}\ \bibnamefont
  {Christiansen}},\ }\href {\doibase 10.1103/PhysRevD.92.094010} {\bibfield
  {journal} {\bibinfo  {journal} {Phys. Rev. D}\ }\textbf {\bibinfo {volume}
  {92}},\ \bibinfo {pages} {094010} (\bibinfo {year} {2015})},\ \Eprint
  {http://arxiv.org/abs/1507.02091} {arXiv:1507.02091 [hep-ph]} \BibitemShut
  {NoStop}%
\bibitem [{\citenamefont {Bierlich}\ \emph {et~al.}(2024)\citenamefont
  {Bierlich}, \citenamefont {Gustafson}, \citenamefont {L{\"o}nnblad},\ and\
  \citenamefont {Shah}}]{Bierlich:2023okq}%
  \BibitemOpen
  \bibfield  {author} {\bibinfo {author} {\bibfnamefont {C.}~\bibnamefont
  {Bierlich}}, \bibinfo {author} {\bibfnamefont {G.}~\bibnamefont {Gustafson}},
  \bibinfo {author} {\bibfnamefont {L.}~\bibnamefont {L{\"o}nnblad}}, \ and\
  \bibinfo {author} {\bibfnamefont {H.}~\bibnamefont {Shah}},\ }\href {\doibase
  10.1140/epjc/s10052-024-12593-0} {\bibfield  {journal} {\bibinfo  {journal}
  {Eur. Phys. J. C}\ }\textbf {\bibinfo {volume} {84}},\ \bibinfo {pages} {231}
  (\bibinfo {year} {2024})},\ \Eprint {http://arxiv.org/abs/2309.12452}
  {arXiv:2309.12452 [hep-ph]} \BibitemShut {NoStop}%
\bibitem [{\citenamefont {Zhang}\ \emph {et~al.}(2025)\citenamefont {Zhang},
  \citenamefont {Peng}, \citenamefont {Peng},\ and\ \citenamefont
  {Zheng}}]{Zhang:2025pqu}%
  \BibitemOpen
  \bibfield  {author} {\bibinfo {author} {\bibfnamefont {A.-G.}\ \bibnamefont
  {Zhang}}, \bibinfo {author} {\bibfnamefont {X.-Y.}\ \bibnamefont {Peng}},
  \bibinfo {author} {\bibfnamefont {X.}~\bibnamefont {Peng}}, \ and\ \bibinfo
  {author} {\bibfnamefont {L.}~\bibnamefont {Zheng}},\ }\href {\doibase
  10.1007/s41365-025-01728-x} {\bibfield  {journal} {\bibinfo  {journal} {Nucl.
  Sci. Tech.}\ }\textbf {\bibinfo {volume} {36}},\ \bibinfo {pages} {134}
  (\bibinfo {year} {2025})},\ \Eprint {http://arxiv.org/abs/2503.10157}
  {arXiv:2503.10157 [hep-ph]} \BibitemShut {NoStop}%
\bibitem [{\citenamefont {Zheng}\ \emph {et~al.}(2024)\citenamefont {Zheng},
  \citenamefont {Liu}, \citenamefont {Lin}, \citenamefont {Shou},\ and\
  \citenamefont {Yin}}]{Zheng:2024xyv}%
  \BibitemOpen
  \bibfield  {author} {\bibinfo {author} {\bibfnamefont {L.}~\bibnamefont
  {Zheng}}, \bibinfo {author} {\bibfnamefont {L.}~\bibnamefont {Liu}}, \bibinfo
  {author} {\bibfnamefont {Z.-W.}\ \bibnamefont {Lin}}, \bibinfo {author}
  {\bibfnamefont {Q.-Y.}\ \bibnamefont {Shou}}, \ and\ \bibinfo {author}
  {\bibfnamefont {Z.-B.}\ \bibnamefont {Yin}},\ }\href {\doibase
  10.1140/epjc/s10052-024-13378-1} {\bibfield  {journal} {\bibinfo  {journal}
  {Eur. Phys. J. C}\ }\textbf {\bibinfo {volume} {84}},\ \bibinfo {pages}
  {1029} (\bibinfo {year} {2024})},\ \Eprint {http://arxiv.org/abs/2404.18829}
  {arXiv:2404.18829 [nucl-th]} \BibitemShut {NoStop}%
\bibitem [{\citenamefont {Zheng}\ \emph {et~al.}(2020)\citenamefont {Zheng},
  \citenamefont {Zhang}, \citenamefont {Shi},\ and\ \citenamefont
  {Lin}}]{Zheng:2019alz}%
  \BibitemOpen
  \bibfield  {author} {\bibinfo {author} {\bibfnamefont {L.}~\bibnamefont
  {Zheng}}, \bibinfo {author} {\bibfnamefont {C.}~\bibnamefont {Zhang}},
  \bibinfo {author} {\bibfnamefont {S.~S.}\ \bibnamefont {Shi}}, \ and\
  \bibinfo {author} {\bibfnamefont {Z.~W.}\ \bibnamefont {Lin}},\ }\href
  {\doibase 10.1103/PhysRevC.101.034905} {\bibfield  {journal} {\bibinfo
  {journal} {Phys. Rev. C}\ }\textbf {\bibinfo {volume} {101}},\ \bibinfo
  {pages} {034905} (\bibinfo {year} {2020})},\ \Eprint
  {http://arxiv.org/abs/1909.07191} {arXiv:1909.07191 [nucl-th]} \BibitemShut
  {NoStop}%
\bibitem [{\citenamefont {Lin}\ and\ \citenamefont
  {Zheng}(2021)}]{Lin:2021mdn}%
  \BibitemOpen
  \bibfield  {author} {\bibinfo {author} {\bibfnamefont {Z.-W.}\ \bibnamefont
  {Lin}}\ and\ \bibinfo {author} {\bibfnamefont {L.}~\bibnamefont {Zheng}},\
  }\href {\doibase 10.1007/s41365-021-00944-5} {\bibfield  {journal} {\bibinfo
  {journal} {Nucl. Sci. Tech.}\ }\textbf {\bibinfo {volume} {32}},\ \bibinfo
  {pages} {113} (\bibinfo {year} {2021})},\ \Eprint
  {http://arxiv.org/abs/2110.02989} {arXiv:2110.02989 [nucl-th]} \BibitemShut
  {NoStop}%
\bibitem [{\citenamefont {Acharya}\ \emph
  {et~al.}(2023{\natexlab{b}})\citenamefont {Acharya} \emph
  {et~al.}}]{ALICE:2023brx}%
  \BibitemOpen
  \bibfield  {author} {\bibinfo {author} {\bibfnamefont {S.}~\bibnamefont
  {Acharya}} \emph {et~al.} (\bibinfo {collaboration} {ALICE}),\ }\href
  {\doibase 10.1007/JHEP10(2023)092} {\bibfield  {journal} {\bibinfo  {journal}
  {JHEP}\ }\textbf {\bibinfo {volume} {10}},\ \bibinfo {pages} {092} (\bibinfo
  {year} {2023}{\natexlab{b}})},\ \Eprint {http://arxiv.org/abs/2302.07783}
  {arXiv:2302.07783 [nucl-ex]} \BibitemShut {NoStop}%
\bibitem [{\citenamefont {Acharya}\ \emph
  {et~al.}(2021{\natexlab{a}})\citenamefont {Acharya} \emph
  {et~al.}}]{ALICE:2021mgk}%
  \BibitemOpen
  \bibfield  {author} {\bibinfo {author} {\bibfnamefont {S.}~\bibnamefont
  {Acharya}} \emph {et~al.} (\bibinfo {collaboration} {ALICE}),\ }\href
  {\doibase 10.1007/JHEP05(2021)220} {\bibfield  {journal} {\bibinfo  {journal}
  {JHEP}\ }\textbf {\bibinfo {volume} {05}},\ \bibinfo {pages} {220} (\bibinfo
  {year} {2021}{\natexlab{a}})},\ \Eprint {http://arxiv.org/abs/2102.13601}
  {arXiv:2102.13601 [nucl-ex]} \BibitemShut {NoStop}%
\bibitem [{\citenamefont {Acharya}\ \emph {et~al.}(2024)\citenamefont {Acharya}
  \emph {et~al.}}]{ALICE:2024xln}%
  \BibitemOpen
  \bibfield  {author} {\bibinfo {author} {\bibfnamefont {S.}~\bibnamefont
  {Acharya}} \emph {et~al.} (\bibinfo {collaboration} {ALICE}),\ }\href
  {\doibase 10.1007/JHEP10(2024)110} {\bibfield  {journal} {\bibinfo  {journal}
  {JHEP}\ }\textbf {\bibinfo {volume} {10}},\ \bibinfo {pages} {110} (\bibinfo
  {year} {2024})},\ \Eprint {http://arxiv.org/abs/2402.16417} {arXiv:2402.16417
  [hep-ex]} \BibitemShut {NoStop}%
\bibitem [{\citenamefont {Acharya}\ \emph
  {et~al.}(2023{\natexlab{c}})\citenamefont {Acharya} \emph
  {et~al.}}]{ALICE:2023wbx}%
  \BibitemOpen
  \bibfield  {author} {\bibinfo {author} {\bibfnamefont {S.}~\bibnamefont
  {Acharya}} \emph {et~al.} (\bibinfo {collaboration} {ALICE}),\ }\href
  {\doibase 10.1103/PhysRevD.108.112003} {\bibfield  {journal} {\bibinfo
  {journal} {Phys. Rev. D}\ }\textbf {\bibinfo {volume} {108}},\ \bibinfo
  {pages} {112003} (\bibinfo {year} {2023}{\natexlab{c}})},\ \Eprint
  {http://arxiv.org/abs/2308.04873} {arXiv:2308.04873 [hep-ex]} \BibitemShut
  {NoStop}%
\bibitem [{\citenamefont {Bai}\ \emph {et~al.}(2024)\citenamefont {Bai},
  \citenamefont {Li}, \citenamefont {Zhang}, \citenamefont {Situ},\ and\
  \citenamefont {Chen}}]{Bai:2024pxk}%
  \BibitemOpen
  \bibfield  {author} {\bibinfo {author} {\bibfnamefont {X.}~\bibnamefont
  {Bai}}, \bibinfo {author} {\bibfnamefont {G.}~\bibnamefont {Li}}, \bibinfo
  {author} {\bibfnamefont {Y.}~\bibnamefont {Zhang}}, \bibinfo {author}
  {\bibfnamefont {Q.}~\bibnamefont {Situ}}, \ and\ \bibinfo {author}
  {\bibfnamefont {X.}~\bibnamefont {Chen}},\ }\href {\doibase
  10.1007/JHEP11(2024)018} {\bibfield  {journal} {\bibinfo  {journal} {JHEP}\
  }\textbf {\bibinfo {volume} {11}},\ \bibinfo {pages} {018} (\bibinfo {year}
  {2024})},\ \Eprint {http://arxiv.org/abs/2405.01444} {arXiv:2405.01444
  [nucl-ex]} \BibitemShut {NoStop}%
\bibitem [{\citenamefont {Aaij}\ \emph {et~al.}(2024)\citenamefont {Aaij} \emph
  {et~al.}}]{LHCb:2023wbo}%
  \BibitemOpen
  \bibfield  {author} {\bibinfo {author} {\bibfnamefont {R.}~\bibnamefont
  {Aaij}} \emph {et~al.} (\bibinfo {collaboration} {LHCb}),\ }\href {\doibase
  10.1103/PhysRevLett.132.081901} {\bibfield  {journal} {\bibinfo  {journal}
  {Phys. Rev. Lett.}\ }\textbf {\bibinfo {volume} {132}},\ \bibinfo {pages}
  {081901} (\bibinfo {year} {2024})},\ \Eprint
  {http://arxiv.org/abs/2310.12278} {arXiv:2310.12278 [hep-ex]} \BibitemShut
  {NoStop}%
\bibitem [{\citenamefont {Aaij}\ \emph
  {et~al.}(2017{\natexlab{a}})\citenamefont {Aaij} \emph
  {et~al.}}]{LHCb:2016qpe}%
  \BibitemOpen
  \bibfield  {author} {\bibinfo {author} {\bibfnamefont {R.}~\bibnamefont
  {Aaij}} \emph {et~al.} (\bibinfo {collaboration} {LHCb}),\ }\href {\doibase
  10.1103/PhysRevLett.118.052002} {\bibfield  {journal} {\bibinfo  {journal}
  {Phys. Rev. Lett.}\ }\textbf {\bibinfo {volume} {118}},\ \bibinfo {pages}
  {052002} (\bibinfo {year} {2017}{\natexlab{a}})},\ \bibinfo {note} {[Erratum:
  Phys.Rev.Lett. 119, 169901 (2017)]},\ \Eprint
  {http://arxiv.org/abs/1612.05140} {arXiv:1612.05140 [hep-ex]} \BibitemShut
  {NoStop}%
\bibitem [{\citenamefont {Aaij}\ \emph
  {et~al.}(2017{\natexlab{b}})\citenamefont {Aaij} \emph
  {et~al.}}]{LHCb:2017vec}%
  \BibitemOpen
  \bibfield  {author} {\bibinfo {author} {\bibfnamefont {R.}~\bibnamefont
  {Aaij}} \emph {et~al.} (\bibinfo {collaboration} {LHCb}),\ }\href {\doibase
  10.1007/JHEP12(2017)026} {\bibfield  {journal} {\bibinfo  {journal} {JHEP}\
  }\textbf {\bibinfo {volume} {12}},\ \bibinfo {pages} {026} (\bibinfo {year}
  {2017}{\natexlab{b}})},\ \Eprint {http://arxiv.org/abs/1710.04921}
  {arXiv:1710.04921 [hep-ex]} \BibitemShut {NoStop}%
\bibitem [{\citenamefont {Abdallah}\ \emph {et~al.}(2011)\citenamefont
  {Abdallah} \emph {et~al.}}]{DELPHI:2011aa}%
  \BibitemOpen
  \bibfield  {author} {\bibinfo {author} {\bibfnamefont {J.}~\bibnamefont
  {Abdallah}} \emph {et~al.} (\bibinfo {collaboration} {DELPHI}),\ }\href
  {\doibase 10.1140/epjc/s10052-011-1557-x} {\bibfield  {journal} {\bibinfo
  {journal} {Eur. Phys. J. C}\ }\textbf {\bibinfo {volume} {71}},\ \bibinfo
  {pages} {1557} (\bibinfo {year} {2011})},\ \Eprint
  {http://arxiv.org/abs/1102.4748} {arXiv:1102.4748 [hep-ex]} \BibitemShut
  {NoStop}%
\bibitem [{\citenamefont {Bierlich}\ \emph {et~al.}(2022)\citenamefont
  {Bierlich} \emph {et~al.}}]{Bierlich:2022pfr}%
  \BibitemOpen
  \bibfield  {author} {\bibinfo {author} {\bibfnamefont {C.}~\bibnamefont
  {Bierlich}} \emph {et~al.},\ }\href {\doibase 10.21468/SciPostPhysCodeb.8}
  {\bibfield  {journal} {\bibinfo  {journal} {SciPost Phys. Codeb.}\ }\textbf
  {\bibinfo {volume} {2022}},\ \bibinfo {pages} {8} (\bibinfo {year} {2022})},\
  \Eprint {http://arxiv.org/abs/2203.11601} {arXiv:2203.11601 [hep-ph]}
  \BibitemShut {NoStop}%
\bibitem [{\citenamefont {Lin}\ and\ \citenamefont {Ko}(2002)}]{Lin:2001zk}%
  \BibitemOpen
  \bibfield  {author} {\bibinfo {author} {\bibfnamefont {Z.-w.}\ \bibnamefont
  {Lin}}\ and\ \bibinfo {author} {\bibfnamefont {C.~M.}\ \bibnamefont {Ko}},\
  }\href {\doibase 10.1103/PhysRevC.65.034904} {\bibfield  {journal} {\bibinfo
  {journal} {Phys. Rev. C}\ }\textbf {\bibinfo {volume} {65}},\ \bibinfo
  {pages} {034904} (\bibinfo {year} {2002})},\ \Eprint
  {http://arxiv.org/abs/nucl-th/0108039} {arXiv:nucl-th/0108039} \BibitemShut
  {NoStop}%
\bibitem [{\citenamefont {Lin}\ \emph {et~al.}(2002)\citenamefont {Lin},
  \citenamefont {Ko},\ and\ \citenamefont {Pal}}]{Lin:2002gc}%
  \BibitemOpen
  \bibfield  {author} {\bibinfo {author} {\bibfnamefont {Z.-w.}\ \bibnamefont
  {Lin}}, \bibinfo {author} {\bibfnamefont {C.~M.}\ \bibnamefont {Ko}}, \ and\
  \bibinfo {author} {\bibfnamefont {S.}~\bibnamefont {Pal}},\ }\href {\doibase
  10.1103/PhysRevLett.89.152301} {\bibfield  {journal} {\bibinfo  {journal}
  {Phys. Rev. Lett.}\ }\textbf {\bibinfo {volume} {89}},\ \bibinfo {pages}
  {152301} (\bibinfo {year} {2002})},\ \Eprint
  {http://arxiv.org/abs/nucl-th/0204054} {arXiv:nucl-th/0204054} \BibitemShut
  {NoStop}%
\bibitem [{\citenamefont {Lin}\ and\ \citenamefont {Ko}(2004)}]{Lin:2003iq}%
  \BibitemOpen
  \bibfield  {author} {\bibinfo {author} {\bibfnamefont {Z.-w.}\ \bibnamefont
  {Lin}}\ and\ \bibinfo {author} {\bibfnamefont {C.~M.}\ \bibnamefont {Ko}},\
  }\href {\doibase 10.1088/0954-3899/30/1/031} {\bibfield  {journal} {\bibinfo
  {journal} {J. Phys. G}\ }\textbf {\bibinfo {volume} {30}},\ \bibinfo {pages}
  {S263} (\bibinfo {year} {2004})},\ \Eprint
  {http://arxiv.org/abs/nucl-th/0305069} {arXiv:nucl-th/0305069} \BibitemShut
  {NoStop}%
\bibitem [{\citenamefont {Zheng}\ \emph {et~al.}(2021)\citenamefont {Zheng},
  \citenamefont {Zhang}, \citenamefont {Liu}, \citenamefont {Lin},
  \citenamefont {Shou},\ and\ \citenamefont {Yin}}]{Zheng:2021jrr}%
  \BibitemOpen
  \bibfield  {author} {\bibinfo {author} {\bibfnamefont {L.}~\bibnamefont
  {Zheng}}, \bibinfo {author} {\bibfnamefont {G.-H.}\ \bibnamefont {Zhang}},
  \bibinfo {author} {\bibfnamefont {Y.-F.}\ \bibnamefont {Liu}}, \bibinfo
  {author} {\bibfnamefont {Z.-W.}\ \bibnamefont {Lin}}, \bibinfo {author}
  {\bibfnamefont {Q.-Y.}\ \bibnamefont {Shou}}, \ and\ \bibinfo {author}
  {\bibfnamefont {Z.-B.}\ \bibnamefont {Yin}},\ }\href {\doibase
  10.1140/epjc/s10052-021-09527-5} {\bibfield  {journal} {\bibinfo  {journal}
  {Eur. Phys. J. C}\ }\textbf {\bibinfo {volume} {81}},\ \bibinfo {pages} {755}
  (\bibinfo {year} {2021})},\ \Eprint {http://arxiv.org/abs/2104.05998}
  {arXiv:2104.05998 [hep-ph]} \BibitemShut {NoStop}%
\bibitem [{\citenamefont {Zhang}(1998)}]{Zhang:1997ej}%
  \BibitemOpen
  \bibfield  {author} {\bibinfo {author} {\bibfnamefont {B.}~\bibnamefont
  {Zhang}},\ }\href {\doibase 10.1016/S0010-4655(98)00010-1} {\bibfield
  {journal} {\bibinfo  {journal} {Comput. Phys. Commun.}\ }\textbf {\bibinfo
  {volume} {109}},\ \bibinfo {pages} {193} (\bibinfo {year} {1998})},\ \Eprint
  {http://arxiv.org/abs/nucl-th/9709009} {arXiv:nucl-th/9709009} \BibitemShut
  {NoStop}%
\bibitem [{\citenamefont {He}\ and\ \citenamefont {Lin}(2017)}]{He:2017tla}%
  \BibitemOpen
  \bibfield  {author} {\bibinfo {author} {\bibfnamefont {Y.}~\bibnamefont
  {He}}\ and\ \bibinfo {author} {\bibfnamefont {Z.-W.}\ \bibnamefont {Lin}},\
  }\href {\doibase 10.1103/PhysRevC.96.014910} {\bibfield  {journal} {\bibinfo
  {journal} {Phys. Rev. C}\ }\textbf {\bibinfo {volume} {96}},\ \bibinfo
  {pages} {014910} (\bibinfo {year} {2017})},\ \Eprint
  {http://arxiv.org/abs/1703.02673} {arXiv:1703.02673 [nucl-th]} \BibitemShut
  {NoStop}%
\bibitem [{\citenamefont {Li}\ and\ \citenamefont {Ko}(1995)}]{Li:1995pra}%
  \BibitemOpen
  \bibfield  {author} {\bibinfo {author} {\bibfnamefont {B.-A.}\ \bibnamefont
  {Li}}\ and\ \bibinfo {author} {\bibfnamefont {C.~M.}\ \bibnamefont {Ko}},\
  }\href {\doibase 10.1103/PhysRevC.52.2037} {\bibfield  {journal} {\bibinfo
  {journal} {Phys. Rev. C}\ }\textbf {\bibinfo {volume} {52}},\ \bibinfo
  {pages} {2037} (\bibinfo {year} {1995})},\ \Eprint
  {http://arxiv.org/abs/nucl-th/9505016} {arXiv:nucl-th/9505016} \BibitemShut
  {NoStop}%
\bibitem [{\citenamefont {Li}\ \emph {et~al.}(2001)\citenamefont {Li},
  \citenamefont {Sustich}, \citenamefont {Zhang},\ and\ \citenamefont
  {Ko}}]{Li:2001xh}%
  \BibitemOpen
  \bibfield  {author} {\bibinfo {author} {\bibfnamefont {B.}~\bibnamefont
  {Li}}, \bibinfo {author} {\bibfnamefont {A.~T.}\ \bibnamefont {Sustich}},
  \bibinfo {author} {\bibfnamefont {B.}~\bibnamefont {Zhang}}, \ and\ \bibinfo
  {author} {\bibfnamefont {C.~M.}\ \bibnamefont {Ko}},\ }\href {\doibase
  10.1142/S0218301301000575} {\bibfield  {journal} {\bibinfo  {journal} {Int.
  J. Mod. Phys. E}\ }\textbf {\bibinfo {volume} {10}},\ \bibinfo {pages} {267}
  (\bibinfo {year} {2001})}\BibitemShut {NoStop}%
\bibitem [{\citenamefont {Butenschoen}\ \emph {et~al.}(2016)\citenamefont
  {Butenschoen}, \citenamefont {Dehnadi}, \citenamefont {Hoang}, \citenamefont
  {Mateu}, \citenamefont {Preisser},\ and\ \citenamefont
  {Stewart}}]{Butenschoen:2016lpz}%
  \BibitemOpen
  \bibfield  {author} {\bibinfo {author} {\bibfnamefont {M.}~\bibnamefont
  {Butenschoen}}, \bibinfo {author} {\bibfnamefont {B.}~\bibnamefont
  {Dehnadi}}, \bibinfo {author} {\bibfnamefont {A.~H.}\ \bibnamefont {Hoang}},
  \bibinfo {author} {\bibfnamefont {V.}~\bibnamefont {Mateu}}, \bibinfo
  {author} {\bibfnamefont {M.}~\bibnamefont {Preisser}}, \ and\ \bibinfo
  {author} {\bibfnamefont {I.~W.}\ \bibnamefont {Stewart}},\ }\href {\doibase
  10.1103/PhysRevLett.117.232001} {\bibfield  {journal} {\bibinfo  {journal}
  {Phys. Rev. Lett.}\ }\textbf {\bibinfo {volume} {117}},\ \bibinfo {pages}
  {232001} (\bibinfo {year} {2016})},\ \Eprint
  {http://arxiv.org/abs/1608.01318} {arXiv:1608.01318 [hep-ph]} \BibitemShut
  {NoStop}%
\bibitem [{\citenamefont {Nason}\ \emph {et~al.}(1999)\citenamefont {Nason}
  \emph {et~al.}}]{Nason:1999ta}%
  \BibitemOpen
  \bibfield  {author} {\bibinfo {author} {\bibfnamefont {P.}~\bibnamefont
  {Nason}} \emph {et~al.},\ }in\ \href {\doibase 10.5170/CERN-2000-004.231}
  {\emph {\bibinfo {booktitle} {{Workshop on Standard Model Physics (and more)
  at the LHC (First Plenary Meeting)}}}}\ (\bibinfo {year} {1999})\ pp.\
  \bibinfo {pages} {231--304},\ \Eprint {http://arxiv.org/abs/hep-ph/0003142}
  {arXiv:hep-ph/0003142} \BibitemShut {NoStop}%
\bibitem [{\citenamefont {Zhang}\ \emph {et~al.}(2019)\citenamefont {Zhang},
  \citenamefont {Zheng}, \citenamefont {Liu}, \citenamefont {Shi},\ and\
  \citenamefont {Lin}}]{Zhang:2019utb}%
  \BibitemOpen
  \bibfield  {author} {\bibinfo {author} {\bibfnamefont {C.}~\bibnamefont
  {Zhang}}, \bibinfo {author} {\bibfnamefont {L.}~\bibnamefont {Zheng}},
  \bibinfo {author} {\bibfnamefont {F.}~\bibnamefont {Liu}}, \bibinfo {author}
  {\bibfnamefont {S.}~\bibnamefont {Shi}}, \ and\ \bibinfo {author}
  {\bibfnamefont {Z.-W.}\ \bibnamefont {Lin}},\ }\href {\doibase
  10.1103/PhysRevC.99.064906} {\bibfield  {journal} {\bibinfo  {journal} {Phys.
  Rev. C}\ }\textbf {\bibinfo {volume} {99}},\ \bibinfo {pages} {064906}
  (\bibinfo {year} {2019})},\ \Eprint {http://arxiv.org/abs/1903.03292}
  {arXiv:1903.03292 [nucl-th]} \BibitemShut {NoStop}%
\bibitem [{\citenamefont {Acharya}\ \emph
  {et~al.}(2021{\natexlab{b}})\citenamefont {Acharya} \emph
  {et~al.}}]{ALICE:2020wfu}%
  \BibitemOpen
  \bibfield  {author} {\bibinfo {author} {\bibfnamefont {S.}~\bibnamefont
  {Acharya}} \emph {et~al.} (\bibinfo {collaboration} {ALICE}),\ }\href
  {\doibase 10.1103/PhysRevLett.127.202301} {\bibfield  {journal} {\bibinfo
  {journal} {Phys. Rev. Lett.}\ }\textbf {\bibinfo {volume} {127}},\ \bibinfo
  {pages} {202301} (\bibinfo {year} {2021}{\natexlab{b}})},\ \Eprint
  {http://arxiv.org/abs/2011.06078} {arXiv:2011.06078 [nucl-ex]} \BibitemShut
  {NoStop}%
\bibitem [{\citenamefont {Amhis}\ \emph {et~al.}(2021)\citenamefont {Amhis}
  \emph {et~al.}}]{HFLAV:2019otj}%
  \BibitemOpen
  \bibfield  {author} {\bibinfo {author} {\bibfnamefont {Y.~S.}\ \bibnamefont
  {Amhis}} \emph {et~al.} (\bibinfo {collaboration} {HFLAV}),\ }\href {\doibase
  10.1140/epjc/s10052-020-8156-7} {\bibfield  {journal} {\bibinfo  {journal}
  {Eur. Phys. J. C}\ }\textbf {\bibinfo {volume} {81}},\ \bibinfo {pages} {226}
  (\bibinfo {year} {2021})},\ \Eprint {http://arxiv.org/abs/1909.12524}
  {arXiv:1909.12524 [hep-ex]} \BibitemShut {NoStop}%
\bibitem [{\citenamefont {Zhao}\ \emph
  {et~al.}(2024{\natexlab{c}})\citenamefont {Zhao}, \citenamefont {Aichelin},
  \citenamefont {Gossiaux},\ and\ \citenamefont {Werner}}]{Zhao:2023ucp}%
  \BibitemOpen
  \bibfield  {author} {\bibinfo {author} {\bibfnamefont {J.}~\bibnamefont
  {Zhao}}, \bibinfo {author} {\bibfnamefont {J.}~\bibnamefont {Aichelin}},
  \bibinfo {author} {\bibfnamefont {P.~B.}\ \bibnamefont {Gossiaux}}, \ and\
  \bibinfo {author} {\bibfnamefont {K.}~\bibnamefont {Werner}},\ }\href
  {\doibase 10.1103/PhysRevD.109.054011} {\bibfield  {journal} {\bibinfo
  {journal} {Phys. Rev. D}\ }\textbf {\bibinfo {volume} {109}},\ \bibinfo
  {pages} {054011} (\bibinfo {year} {2024}{\natexlab{c}})},\ \Eprint
  {http://arxiv.org/abs/2310.08684} {arXiv:2310.08684 [hep-ph]} \BibitemShut
  {NoStop}%
\bibitem [{\citenamefont {Acharya}\ \emph
  {et~al.}(2022{\natexlab{a}})\citenamefont {Acharya} \emph
  {et~al.}}]{ALICE:2021npz}%
  \BibitemOpen
  \bibfield  {author} {\bibinfo {author} {\bibfnamefont {S.}~\bibnamefont
  {Acharya}} \emph {et~al.} (\bibinfo {collaboration} {ALICE}),\ }\href
  {\doibase 10.1016/j.physletb.2022.137065} {\bibfield  {journal} {\bibinfo
  {journal} {Phys. Lett. B}\ }\textbf {\bibinfo {volume} {829}},\ \bibinfo
  {pages} {137065} (\bibinfo {year} {2022}{\natexlab{a}})},\ \Eprint
  {http://arxiv.org/abs/2111.11948} {arXiv:2111.11948 [nucl-ex]} \BibitemShut
  {NoStop}%
\bibitem [{\citenamefont {Acharya}\ \emph
  {et~al.}(2022{\natexlab{b}})\citenamefont {Acharya} \emph
  {et~al.}}]{ALICE:2021rzj}%
  \BibitemOpen
  \bibfield  {author} {\bibinfo {author} {\bibfnamefont {S.}~\bibnamefont
  {Acharya}} \emph {et~al.} (\bibinfo {collaboration} {ALICE}),\ }\href
  {\doibase 10.1103/PhysRevLett.128.012001} {\bibfield  {journal} {\bibinfo
  {journal} {Phys. Rev. Lett.}\ }\textbf {\bibinfo {volume} {128}},\ \bibinfo
  {pages} {012001} (\bibinfo {year} {2022}{\natexlab{b}})},\ \Eprint
  {http://arxiv.org/abs/2106.08278} {arXiv:2106.08278 [hep-ex]} \BibitemShut
  {NoStop}%
\bibitem [{\citenamefont {Cacciari}\ \emph {et~al.}(2012)\citenamefont
  {Cacciari}, \citenamefont {Frixione}, \citenamefont {Houdeau}, \citenamefont
  {Mangano}, \citenamefont {Nason},\ and\ \citenamefont
  {Ridolfi}}]{Cacciari:2012ny}%
  \BibitemOpen
  \bibfield  {author} {\bibinfo {author} {\bibfnamefont {M.}~\bibnamefont
  {Cacciari}}, \bibinfo {author} {\bibfnamefont {S.}~\bibnamefont {Frixione}},
  \bibinfo {author} {\bibfnamefont {N.}~\bibnamefont {Houdeau}}, \bibinfo
  {author} {\bibfnamefont {M.~L.}\ \bibnamefont {Mangano}}, \bibinfo {author}
  {\bibfnamefont {P.}~\bibnamefont {Nason}}, \ and\ \bibinfo {author}
  {\bibfnamefont {G.}~\bibnamefont {Ridolfi}},\ }\href {\doibase
  10.1007/JHEP10(2012)137} {\bibfield  {journal} {\bibinfo  {journal} {JHEP}\
  }\textbf {\bibinfo {volume} {10}},\ \bibinfo {pages} {137} (\bibinfo {year}
  {2012})},\ \Eprint {http://arxiv.org/abs/1205.6344} {arXiv:1205.6344
  [hep-ph]} \BibitemShut {NoStop}%
\end{thebibliography}%

\appendix 

\section{Comparing PYTHIA8 calculations with $m_b$ variation to FONLL} \label{sec:appendix2}

In this appendix, we provide a comparative study between beauty quark production cross sections calculated using a modified $m_b$ mass parameter in PYTHIA8 and predictions from the FONLL framework. The modification of the beauty quark mass $m_b$ is applied only during the initial production stage in PYTHIA to regulate the total cross section. It is important to emphasize that the FONLL calculation~\cite{Cacciari:2012ny} provides a rigorous theoretical benchmark, as it systematically handles next to leading order corrections and the resummation of large logarithm effects which are inherently absent in the leading order event generator of PYTHIA8.

\begin{figure}[thbp]
	\begin{center}
		\includegraphics[width=0.48\textwidth]{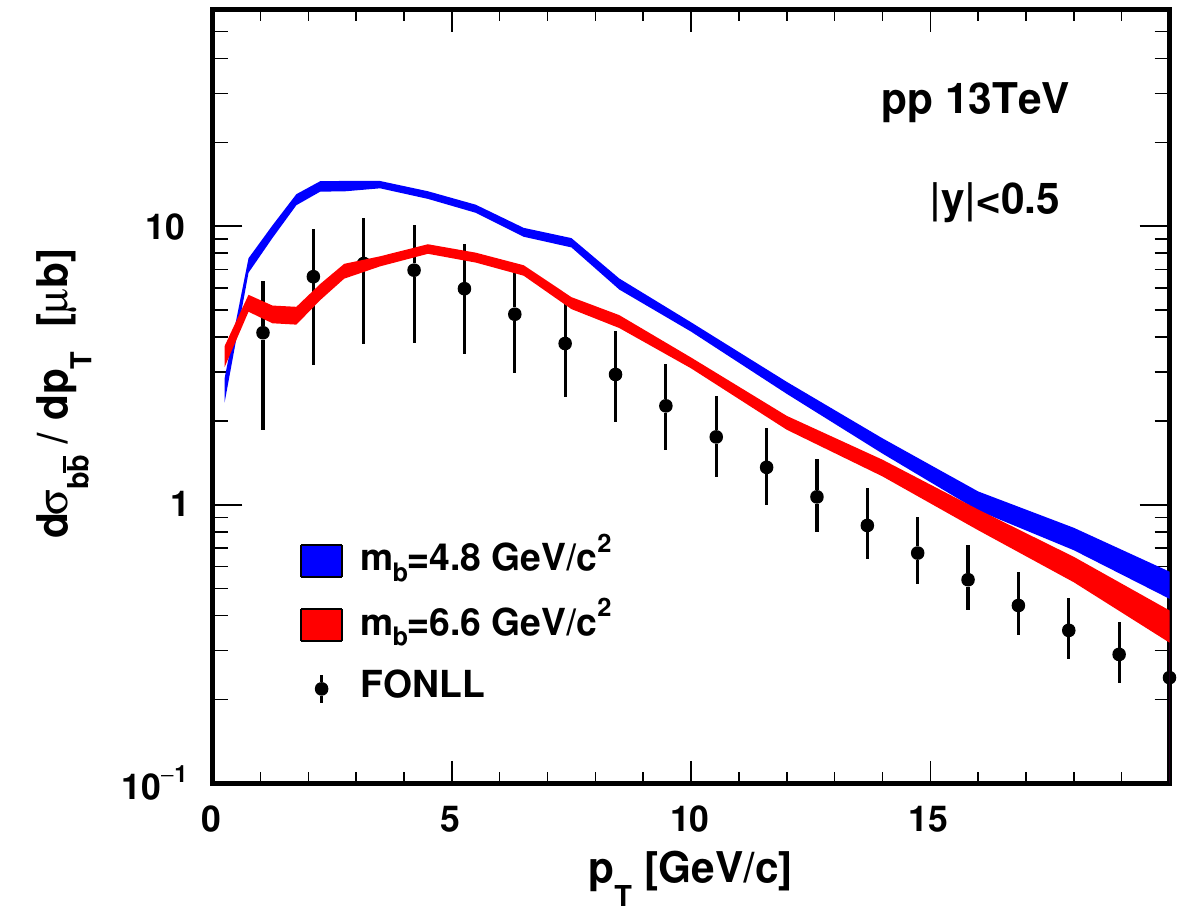}
		\caption{The $p_T$ differential spectra of beauty quarks at mid-rapidity from PYTHIA with different $m_b$, compared with FONLL predictions~\cite{Cacciari:2012ny}.}
		\label{fig:withFONLL}
	\end{center}
\end{figure}

Figure.~\ref{fig:withFONLL} shows the $p_T$ spectra of beauty quarks at mid-rapidity obtained from PYTHIA8 with different values of $m_b$, compared to the FONLL calculation~\cite{Cacciari:2012ny}. The default PYTHIA8 simulation using $m_b=4.8$ GeV$/c$ tends to significantly overestimate the cross section compared to FONLL. Increasing $m_b$ significantly suppresses the yield at low $p_T$ and develops a dip-like structure due to phase space constraints near threshold. The leading order perturbative QCD approach employed in PYTHIA8 tends to overestimate the total cross section and yield harder $p_T$ spectra relative to FONLL, which systematically incorporates higher order corrections and resummation effects. The chosen $m_b$ value brings the PYTHIA results closer to the FONLL baseline, thereby improving agreement with experimental data. This tuning effectively mimics the higher order suppression mechanisms absent in the leading order PYTHIA calculation.

\section{Analyzing multiplicity dependence for non-prompt to prompt charm hadron ratios} \label{sec:appendix}

\begin{figure*}[htbp]
	\begin{center}
		\includegraphics[width=1.0\textwidth]{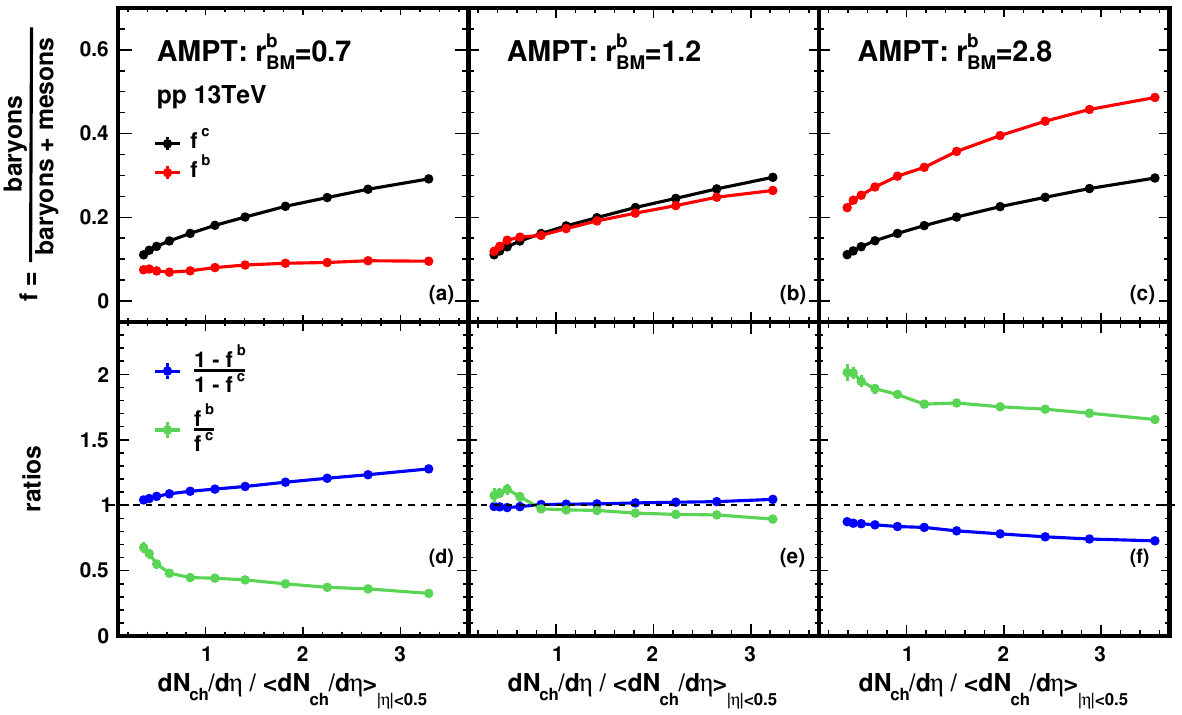}
		\caption{Top row: Multiplicity dependence of $f$ in $pp$ collisions at $\sqrt{s}=13$ TeV in full rapidity space. The black and red represent hadron flavor is charm and beauty ($f^c$ and $f^b$) respectively, with all hadron decay processes turned off. Bottom row: Multiplicity dependence of $(1-f^b)/(1-f^c)$ and $f^b/f^c$, represented by the blue and green markers. From left to right, the panels correspond to $r_{BM}^{b}$ values in AMPT of 0.7, 1.2, and 2.8, respectively.}                 
		\label{fig:fbaryons}
	\end{center}
\end{figure*}

\begin{figure*}[htbp]
	\begin{center}
		\includegraphics[width=1.0\textwidth]{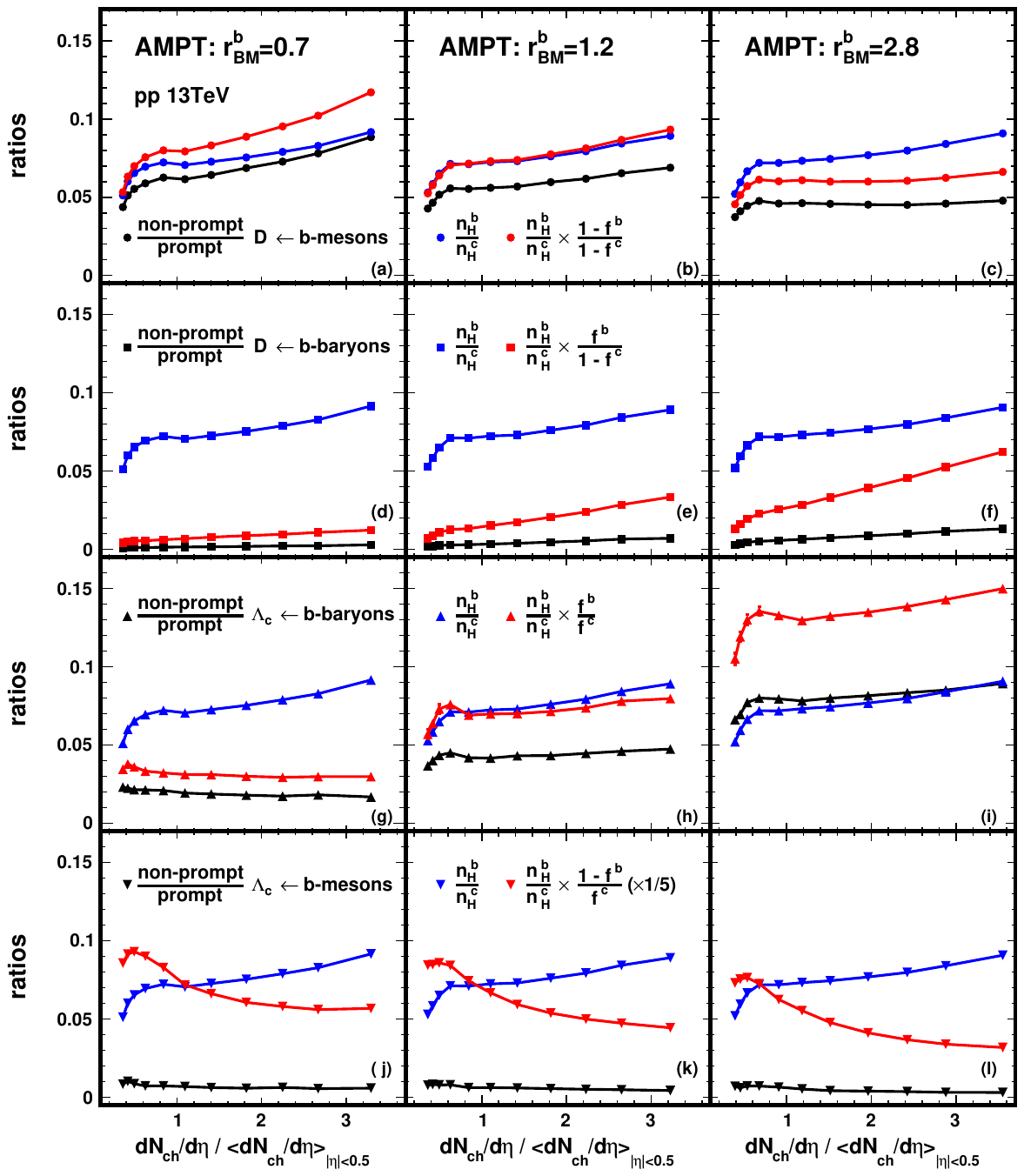}
		\caption{The multiplicity dependence of $\frac{non-prommpt}{prompt}$ (black), $\frac{n_H^b}{n_H^c}$ (blue) and $\frac{n_H^b}{n_H^c}\times$hadronization-fraction (red). From left to right, the panels show AMPT results for $r_{BM}^{b}=$ 0.7, 1.2, and 2.8. From top to bottom, the panels represent the following studied objects: $c$-meson from $b$-meson decays, $c$-meson from $b$-baryon decays, $b$-baryon from $b$-baryon decays, and $b$-baryon from $b$-meson decays, distinguished by different markers respectively.}
		\label{fig:simulate}
	\end{center}
\end{figure*}

To understand the multiplicity dependence of non-prompt to prompt charm hadron production ratios, we divide the analysis of this ratio into three factors: heavy-quark production, hadronization, and the subsequent weak decays of beauty hadrons. The beauty hadron decay process is expected to be largely insensitive to the event multiplicity and can be regarded as a constant determined by the branching ratio. Consequently, the observed multiplicity dependence of non-prompt to prompt production primarily reflects the interplay between the initial heavy quark production and the hadronization dynamics for beauty and charm. For non-prompt charm mesons these dependences can be factorized as follows:
\begin{equation}\label{eq:meson}
\begin{split}
\frac{NP}{P}(M^c) 
&= \frac{n_H^b}{n_H^c} \times \frac{1 - f^b}{1 - f^c} \times BR(M^b\rightarrow M^c) \\
&\quad + \frac{n_H^b}{n_H^c} \times \frac{f^b}{1 - f^c} \times BR(B^b\rightarrow M^c),
\end{split}
\end{equation}
and for non-prompt charm baryons,
\begin{equation}\label{eq:baryon}
\begin{split}
\frac{NP}{P}(B^c)
&= \frac{n_H^b}{n_H^c} \times \frac{f^b}{f^c} \times BR(B^b\rightarrow B^c) \\
&\quad + \frac{n_H^b}{n_H^c} \times \frac{1 - f^b}{f^c} \times BR(M^b\rightarrow B^c).
\end{split}
\end{equation}
The superscripts $c$ or $b$ denote charm or beauty flavor. $n_H^b$ and $n_H^c$ represents the number of beauty and charm hadrons right after hadronization and before weak decay process accordingly. $M$ and $B$ represent the meson and baryon states with different flavor. $f$ is the fraction of baryons among hadrons for each flavor without decay, while the meson fraction is $1-f$. The branching ratios $BR$ specify the decay probabilities from beauty baryons ($B^b$) and mesons ($M^b$) into specific final charm baryons ($B^c$) and mesons ($M^c$). Since the decay sources of non-prompt charm hadrons can be divided into beauty mesons and baryons, our equation is naturally split into two parts representing the sum of beauty meson and baryon decay channel contributions. Importantly, in Eq.\eqref{eq:meson}, the leading contribution to non-prompt charm mesons arises from the $M^b\rightarrow M^c$ decay channel, while in Eq.\eqref{eq:baryon}, the dominant channel is the $B^b\rightarrow B^c$ decay channel. The baryon fraction $f$ is directly influenced by the coalescence parameter $r_{BM}$ and represents the hadronization dynamics within AMPT. In the following discussion, we will compare AMPT results for three choices of $r_{BM}^{b}$ (0.7, 1.2 and 2.8) to illustrate how quark production and hadronization jointly shape the multiplicity dependence of non-prompt to prompt ratios.  

In this factorized formulation, the initial heavy flavor hadron production ratio $n_H^b/n_H^c$ is independent of $r_{BM}$. We can observe that the multiplicity dependence the baryon hadronization fraction, $f$, is largely driven by the coalescence parameter. The top row of Fig.~\ref{fig:fbaryons} shows the multiplicity dependence of $f$ in $pp$ collisions at $\sqrt{s}=13$ TeV, evaluated in full rapidity space with all hadron decay processes turned off. The black and red curves represent the charm ($f^c$) and beauty ($f^b$) hadrons, respectively. We observe that the baryon fraction increases with charged particle multiplicity, reflecting the enhanced importance of coalescence mechanism at higher particle densities. The charm baryon fraction is unchanged in this comparison as $r_{BM}^c$ is fixed to 1.4 throughout this work. 
The bottom row of Fig.~\ref{fig:fbaryons} present the ratios of beauty to charm baryon fractions and beauty to charm meson fractions. When the coalescence parameter for beauty quarks $r_{BM}^b$ is smaller than that for charm $r_{BM}^c$, the growth of the beauty baryon fraction with multiplicity is weaker than that of charm, while the beauty meson fraction increases more strongly. The opposite behavior occurs if $r_{BM}^b>r_{BM}^c$. When the two parameters are comparable, both the baryon and meson fraction ratios approach unity and exhibit little multiplicity dependence. Thus, the relative multiplicity dependence of baryon and meson sectors is directly sensitive to the relative magnitude of the coalescence parameters for beauty and charm, highlighting the role of flavor dependent hadronization dynamics.

Next, we examine how the initial heavy-flavor production, the hadronization process, and subsequent decays combine to determine the multiplicity dependence of the non-prompt to prompt ratio. Figure~\ref{fig:simulate} shows the multiplicity dependence of three quantities: the non-prompt/prompt ratio (black), the initial hadron production ratio $\frac{n_H^b}{n_H^c}$ (blue), and the product of $\frac{n_H^b}{n_H^c}$ with the hadronization fraction (red) shown in Fig.~\ref{fig:fbaryons}. From left to right, the panels show AMPT results for $r_{BM}^{b}=$ 0.7, 1.2, and 2.8. From top to bottom, the rows display, respectively, charm mesons from beauty meson decay, charm mesons from beauty baryon decay, charm baryons from beauty baryon decay and charm baryons from beauty meson decay. Note that we scale the red curve in the bottom row by a factor $1/5$ for better visibility. The blue curves are nearly identical across all cases and show a mild upward trend, reflecting the fact that the PYTHIA8 initial conditions predict a slightly enhanced beauty quark production probability at higher multiplicities. This indicates differences in the way beauty and charm quarks are produced in PYTHIA8 initial conditions. Beauty production is dominated by harder partonic scatterings, which scale more strongly with event activity, while charm production receives substantial contributions from softer gluon radiation process. As multiplicity increases, the relative enhancement from hard processes and gluon splitting benefits beauty quarks more than charm, leading to the slow rise of $\frac{n_H^b}{n_H^c}$. 
By contrast, the red curves ($\frac{n_H^b}{n_H^c}\times$ hadronization fraction) show strong sensitivity to the choice of $r_{BM}^b$. When $r_{BM}^b$ is significantly smaller than $r_{BM}^c$ , the beauty baryon fraction grows more slowly with multiplicity than the charm baryon fraction, causing the red curves for baryons to rise weakly or even saturate, while the meson channel shows a stronger increase. Conversely, when $r_{BM}^b$ is much larger, the baryon channel steepens considerably with multiplicity, while the meson channel is comparatively suppressed. In the case of $r_{BM}^b$ is similar to $r_{BM}^c$, both baryon and meson sectors evolve with similar slopes, yielding nearly flat ratios between beauty and charm fractions.
The overall downward shift of the black points (non-prompt to prompt ratios) relative to the red curves originates from the branching ratios, which act almost as multiplicative constants smaller than unity. This behavior is consistent with the factorized formulation in Eq.~\eqref{eq:meson} and \eqref{eq:baryon}, where the decay contribution enters multiplicatively through the branching ratios, leaving the multiplicity dependence governed primarily by the production and hadronization terms. Finally, it is worth noting that the contributions from beauty baryon to charm meson and beauty meson to charm baryon decays are comparatively small, as seen in the second and fourth row of Fig.~\ref{fig:simulate}. Accordingly, the associated effects on the multiplicity dependence are weak, and the dominant behavior of the non-prompt/prompt ratios is governed by the primary decay channels. Therefore, the contrast across the three columns of Fig.~\ref{fig:simulate} highlights the central role of the coalescence parameter in controlling the flavor dependent baryon meson competition. Varying $r_{BM}^b$ directly reshapes the multiplicity dependence of non-prompt to prompt ratios. This demonstrates that measurements of multiplicity-dependent heavy-flavor observables are driven primarily by the baryon to baryon and meson to meson decay channels, providing particularly sensitive probes of flavor-dependent coalescence dynamics.

\end{document}